\documentclass[prl,aps,twocolumn,nofootinbib,preprintnumbers,superscriptaddress,preprintnumbers,balancelastpage,longbibliography]{revtex4-2}
\usepackage[utf8]{inputenc}
\usepackage[colorlinks=true,citecolor=blue,linkcolor=blue]{hyperref}
\usepackage[normalem]{ulem}
\usepackage{amsmath,amssymb, mathrsfs,xfrac}
\usepackage{epsfig}
\usepackage{graphicx}               
\usepackage{color}
\usepackage{slashed}
\usepackage{multirow}
\usepackage{placeins}
\usepackage{graphicx}
\usepackage[dvipsnames]{xcolor}
\usepackage{epstopdf}
\usepackage{soul}
\usepackage{tikz}
\usepackage[capitalise, english]{cleveref}
\usepackage{siunitx}
\usepackage{xspace}
\usepackage{booktabs}
\usetikzlibrary{trees}
\usepackage{graphicx}
\usepackage[export]{adjustbox}
\usetikzlibrary{decorations.pathmorphing}
\usetikzlibrary{decorations.markings}
\usepackage{aas_macros}
\usepackage{lettrine}
\usepackage{subfigure}
\usepackage{graphicx}
\usepackage{float}

\usepackage{placeins}
\usepackage{textgreek}
\usepackage{booktabs, tabularx, array}
\newcolumntype{Y}{>{\raggedright\arraybackslash}X}

\LettrineTextFont{\itshape}
\usepackage{comment}

\newcommand\myshade{80}

\colorlet{mylinkcolor}{ForestGreen}

\colorlet{mycitecolor}{Aquamarine}

\colorlet{myurlcolor}{violet}

\hypersetup{
  linkcolor  = mylinkcolor!\myshade!black,
  citecolor  = mycitecolor!\myshade!black,
  urlcolor   = myurlcolor!\myshade!black,
  colorlinks = true
}

\definecolor{jblue}{RGB}{20,50,100}
\definecolor{npurple}{RGB} {153, 51, 204}
\definecolor{wred}{RGB}{217,0,56}
\definecolor{white}{RGB}{255,255,255}

\definecolor{korange}{RGB}{235, 80,  43}
\definecolor{korange2}{RGB}{245, 100,  63}
\definecolor{kyelloworange}{RGB}{255, 210,  110}
\definecolor{kyelloworange2}{RGB}{240, 170,  90}
\definecolor{kred}{RGB}{204,  102, 153}
\definecolor{kpurple}{RGB}{153,  61, 190}
\definecolor{kpurplelight}{RGB}{213,  161, 230}

 \definecolor{tobycolour}{rgb}{.5,.0,.5}

\DeclareSIUnit\year{yr}
\DeclareSIUnit\pc{pc}
\DeclareSIUnit\ergs{ergs}
\DeclareSIUnit\msun{\ensuremath{M_\odot}}
\allowdisplaybreaks

\makeatletter
\providecommand*{\diff}%
  {\command{\lmultau}{\ensuremath{L_\mu-L_\tau}\xspace}
\new@ifnextchar^{\DIfF}{\DIfF^{}}}
\def\DIfF^#1{%
  \mathop{\mathrm{\mathstrut d}}%
    \nolimits^{#1}\gobblespace}
\def\gobblespace{%
  \futurelet\diffarg\opspace}
\def\opspace{%
  \let\DiffSpace\!%
  \ifx\diffarg(%
    \let\DiffSpace\relax
  \else
    \ifx\diffarg[%
      \let\DiffSpace\relax
    \else
        \ifx\diffarg\{%
        \let\DiffSpace\relax
      \fi\fi\fi\DiffSpace}

\usepackage{tikz,xcolor,hyperref}

\definecolor{lime}{HTML}{A6CE39}
\DeclareRobustCommand{\orcidicon}{\hspace{-1mm}
	\begin{tikzpicture}
	\draw[lime, fill=lime] (0,0) 
	circle [radius=0.16] 
	node[white] {{\fontfamily{qag}\selectfont \tiny \,ID}};
	\draw[white, fill=white] (-0.0525,0.095) 
	circle [radius=0.007];
	\end{tikzpicture}
	\hspace{-3mm}
}

\foreach \x in {A, ..., Z}{\expandafter\xdef\csname orcid\x\endcsname{\noexpand\href{https://orcid.org/\csname orcidauthor\x\endcsname}
			{\noexpand\orcidicon}}
}

\usepackage{physics}

\newcommand{\mytitle}{Shining LIMelight into the dark: Probing Dark matter - baryon interactions with Line Intensity Mapping}

\begin{document}

	\title{\mytitle}
	
\author{Durba Ghosh\orcidA{}}
\email{durbaghosh@iisc.ac.in}
\affiliation{Centre for High Energy Physics, Indian Institute of Science, C.\,V.\,Raman Avenue, Bengaluru 560012, India}

\author{Ranjini Mondol\orcidB{}}
\email{ranjini12m@gmail.com, ranjini.mondol@saha.ac.in}
\affiliation{Centre for High Energy Physics, Indian Institute of Science, C.\,V.\,Raman Avenue, Bengaluru 560012, India}
\affiliation{Theory Division, Saha Institute of Nuclear Physics, 1/AF, Bidhannagar, Kolkata 700064, India}

\author{Ranjan Laha\orcidD{}}
\email{ranjanlaha@iisc.ac.in}
\affiliation{Centre for High Energy Physics, Indian Institute of Science, C.\,V.\,Raman Avenue, Bengaluru 560012, India}

	\date{\today}

\begin{abstract}
Line-intensity mapping (LIM) is an emerging technique for probing both the astrophysics of galaxy formation and the cosmological evolution of large-scale structure. We investigate, \textit{for the first time}, its potential as a probe of dark matter–baryon (DM–$b$) interactions using the upcoming Carbon Monoxide Mapping Array Project (COMAP) surveys, COMAP-EoR (Epoch of Reionization) and COMAP-ERA (Expanded Reionization Array). Elastic scattering of dark matter (DM) with baryons, i.e., protons ($p$) and electrons ($e$), exchange momentum and heat with the DM, suppressing structure formation and leaving characteristic signatures in the linear matter power spectrum. These effects propagate to the LIM power spectrum (PS) and voxel intensity distribution (VID). We consider momentum-transfer cross sections $\sigma^{\chi b}_{T,n}=\sigma^{\chi b}_{n}v^n$, with $n=0,-2,$ and $-4$, corresponding to \textit{contact}, \textit{electric dipole-moment-like}, and \textit{Coulomb-like} interactions, respectively. Using Fisher forecasts, we obtain the projected sensitivities to $\sigma^{\chi b}_{n}$ from the PS and VID. Our VID-only forecasts combining the 16 and 30 GHz observed bands yield stronger sensitivities than existing cosmological limits for $n=-2$ and $n=-4$ for sub-GeV DM masses. We also explore the potential of combining the PS and VID to probe these interactions. Our study highlights LIM as a complementary probe of DM models alongside other astrophysical, cosmological, and terrestrial searches.
\end{abstract}
\maketitle
\section{Introduction} 
\label{sec:introduction}

The $\Lambda$ cold dark matter ($\Lambda \rm CDM$) framework posits that $\sim 85\%$ of the matter content of the Universe is composed of DM, whose presence is inferred from its gravitational influence on cosmic structures \cite{de_Martino_2022, Planck:2018vyg, Massey:2010hh, Shin_2022, Mastropietro:2007kr, sofue2020rotationcurvemilkyway, Mastropietro:2007kr,10.1093/mnras/staf1391,Lim:2023lss,Cirelli:2024ssz,1970ApJ...159..379R}. Observations ranging from the cosmic microwave background (CMB) to galaxy dynamics across multiple scales provide compelling evidence for its existence, yet its fundamental nature remains unknown. Viable DM candidates span an enormous mass range of $\sim 90$ orders of magnitude. In the standard paradigm, DM is modeled as a cold, non-relativistic, and collisionless component.  Significant attention has been directed towards scenarios in which DM is realized as a particle, exhibiting non-gravitational interactions with Standard Model particles \cite{Cirelli:2024ssz,Bertone:2016nfn,Strigari:2012acq,Slatyer:2017sev,Lin:2019uvt, Cirelli:2024ssz, Akita:2023yga}. A particularly well-studied possibility involves beyond the Standard Model (BSM) elastic scattering between DM and baryons \cite{Buen-Abad:2021mvc,Ali-Haimoud:2023pbi,Blanco:2023bgz,Gluscevic:2017ywp,Dvorkin:2013cea,He:2025npy,Sigurdson:2004zp,Zhang:2024mmg, McDermott:2010pa, Boehm:2004th,Berlin:2018sjs, Escudero:2018thh}, which induces an exchange of momentum and heat between the two components. This coupling suppresses the growth of  gravitational potential, delays structure formation, and leads to a scale-dependent suppression of the linear matter power spectrum $P_{\rm{m}}(k)$, where $k$ denotes the wavenumber. This affects the HMF, altering the abundance of collapsed  objects \cite{Nguyen:2021cnb,Boddy:2018kfv,Ali-Haimoud:2023pbi,Dvorkin:2013cea,Buen-Abad:2021mvc}. 
\begin{figure}[!htbp]
    \centering
\includegraphics[width=0.9\linewidth]{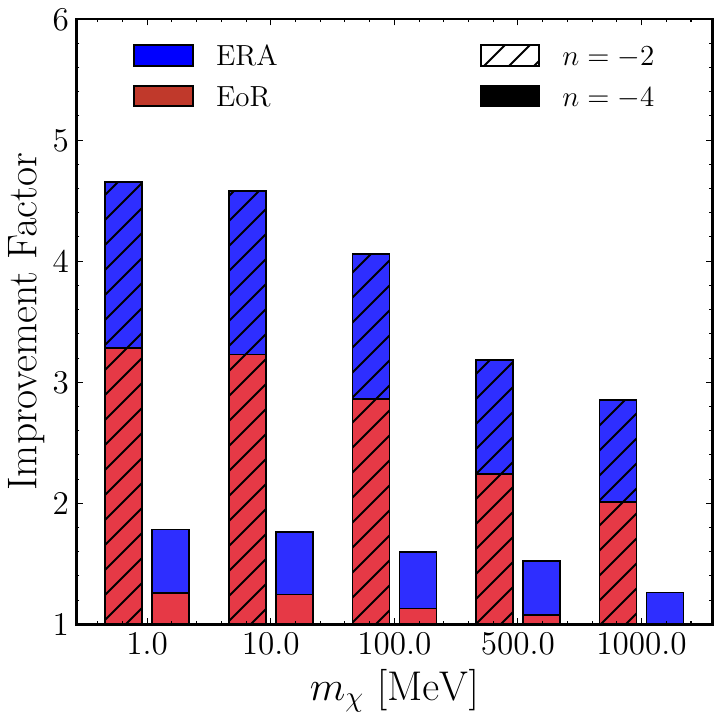}
    \caption{Improvement factor, $\sigma^{\chi p}_{n,\rm best}/\sigma^{\chi p}_{n,\rm LIM}$, as a function of the DM mass $m_{\chi}$, where $\sigma^{\chi p}_{n,\rm LIM}$ and $\sigma^{\chi p}_{n,\rm best}$ denote the 95\% C.L. projected upper limits from LIM experiments and existing best 95\% C.L. upper limits, respectively, on the normalization $\sigma^{\chi p}_{n}$ with velocity dependence $v^n$. We show COMAP-ERA (\textbf{blue}) and COMAP-EoR (\textbf{red}) for $n=-2$ (\textbf{hatched}) and $n=-4$ (\textbf{solid}). In both cases, COMAP-ERA provides the greatest improvement over the current limits. }
    \label{fig:Bar_Plot}
\end{figure}
Considerable efforts have been devoted to identifying signatures of such interactions using a wide range of observational probes, including direct detection experiments \cite{XENON:2016jmt,XENON:2017vdw,CRESST:2019jnq,CDMS:2003yjm}, cosmic-ray measurements \cite{Cappiello:2019qsw,Bringmann:2018cvk}, CMB \cite{Ali-Haimoud:2015pwa,Xu:2018efh,Slatyer:2009yq}, and Milky Way satellite (MWS) counts \cite{Maamari:2020aqz,Esteban:2023xpk,Nadler_2019}, Lyman-$\alpha$ forest \cite{Ooba:2019erm,Bose:2018juc},  21-cm
signal \cite{Rahimieh:2025fsb,Rahimieh:2025lbf, Slatyer:2018aqg, Kovetz:2018zan,Fialkov:2018xre, Munoz:2018pzp, Barkana:2018qrx}, and UV luminosity functions (UVLF) \cite{Das:2025bnr, Lazare:2025gha}.
In this context, LIM has emerged as a promising and complementary observational framework~\cite{Bernal:2022jap,Oxholm:2022mnn,Chang:2026ake,Kovetz:2017agg, Fonseca:2016qqw, Carilli_2011,Gong_2011,Visbal_2011,Visbal_2010,Chang:2007xk, Suginohara:1998ti,10.1093/mnras/188.4.791}.

LIM has recently gained prominence as an efficient approach to rapidly survey large cosmological volumes out to the cosmic dawn. Major observational programs targeting the CMB \cite{Planck:2018vyg} and galaxy surveys (GS) \cite{BOSS:2016wmc} have enabled stringent tests of the $\Lambda \rm{CDM}$ model, yet questions about the nature of DM and the formation and evolution of cosmic structures remain open. GS are limited in detecting and resolving individual discrete objects above a threshold brightness, while the CMB provides an integrated measurement on reionization and can therefore accommodate degeneracies among different ionization histories \cite{Planck:2018vyg}. In contrast, LIM measures the cumulative emission of a specific spectral line from unresolved sources and intergalactic medium (IGM) across large cosmic volumes, making it highly complementary to both GS and CMB observations. Relaxing the need to resolve individual emitters permits efficient, wide-area surveys with relatively low-aperture instruments and modest experimental resources. Recent years have seen a surge of interest in LIM, with a wide range of future and running observations targeting various atomic and molecular lines, from radio to ultraviolet \cite{van_Haarlem_2013, Bandura_2014, DeBoer:2016tnn, MeerKLASS:2017vgf, Keating:2020wlx, Keating:2016pka, COMAP:2021eue, CONCERTO:2020ahk, Gebhardt:2021vfo,Chang:2007xk,Visbal:2008rg,Breysse:2014uia,Lidz_2011,Silva:2014ira,Padmanabhan:2018yul,Gong_2017,Pullen:2013dir,CCAT-Prime:2021lly,Sun:2020mco,vieira2020terahertzintensitymappertim,SPHEREx:2014bgr,Koopmans:2015sua,Newburgh:2016mwi, 10.1117/1.JATIS.7.4.044004, Li:2015gqa, Padmanabhan:2017ate, Lidz_2011, Pullen_2013, Li_2016, Pullen:2017ogs, 10.1093/mnras/stx3265, Silva_2013, Keating:2015qva, Yang:2019eoj, BOSS:2015ids, 10.1117/12.2057207, SPHEREx:2014bgr, Lagache:2017htw, Stacey:2018yqe, Cooray:2016hro, theostmissionconceptstudyteam2018originsspacetelescopeost}, 
alongside substantial advances in theoretical modeling and data analysis.

LIM offers a powerful avenue for investigating both cosmological and astrophysical phenomena. The observed line emission traces the underlying matter density field as a biased tracer of DM fluctuations. It also encodes information about the astrophysical processes governing line emission. Two complementary summary statistics for LIM are the PS and VID \cite{Bernal:2023ovz,Sabla:2024dxz}. The PS is defined as the two-point correlation function of the intensity fluctuations in Fourier space. The VID gives the probability distribution of line intensities across voxels. A \textit{voxel} is a three-dimensional volume element in a survey, with its two transverse dimensions determined by the angular resolution and its line-of-sight (l.o.s.) dimension determined by the spectral resolution~\cite{Chang:2026ake}. The \textit{joint}\footnote{By \textit{joint}, we mean the analysis combining two observables PS and VID.} use of PS and VID helps to reduce parameter degeneracies and improve  resulting sensitivities \cite{COMAP:2018kem, Sabla:2024dxz}. 

In this work, we demonstrate that LIM observables
provide a powerful means to probe DM-$b$ interactions
since it is sensitive to the $P_{\rm{m}}(k)$ and, consequently, the HMF. We focus on COMAP-EoR and COMAP-ERA \cite{COMAP:2021nrp}, the next-generation phases of the COMAP Project\footnote{\url{https://comap.caltech.edu/}}, which observes the CO(1-0) line with rest-frame frequency of 115.27 GHz. Using the Fisher matrix formalism, we forecast their sensitivities to the DM-$b$ cross-section and relevant astrophysical parameters, and explore the degeneracies among them. COMAP-EoR will observe at  central frequencies $\sim 30$ GHz $(z\sim3)$ and $\sim 16$ GHz $(z\sim 6)$, while COMAP-ERA will observe at the same central frequencies and redshift ranges with substantially longer observing time, $t_{obs}$.

Following \cite{Sigurdson:2004zp, Dvorkin:2013cea, Nguyen:2021cnb, Becker:2020hzj, Boddy:2022tyt, Li:2022mdj}, we parametrize the DM-$b$ momentum transfer cross-section as 
$\sigma^{\chi b}_{T, n}(v)=\sigma^{\chi b}_{n}v^n$, where $b\in(e,p)$, $\sigma^{\chi b}_n$ is the normalization of the cross-section and $v$ is the DM–baryon relative velocity, and $n$ depends on the underlying particle physics model. We forecast constraints on $\sigma^{\chi b}_n$ using \textit{combined} \footnote{By \textit{combined} analysis we mean the analysis where both the fre-
quencies 16 GHz and 30 GHz are considered for an observable. This should be distinguished from the \textit{joint} analysis, which combines the PS and VID at a fixed frequency.} PS and \textit{combined} VID analyses for COMAP-EoR and COMAP-ERA, where we incorporate signals from the 30 GHz and 16 GHz bands. We also perform a \textit{joint} analysis using the 30 GHz band. Among the individual observables, the \textit{combined} VID analysis yields the strongest sensitivities, while the \textit{combined} PS analysis provides the weakest. The enhanced constraining power of the VID arises from its greater sensitivity to astrophysical parameters, which helps break degeneracies between astrophysical and cosmological effects, as well as from its ability to capture non-Gaussian information. The \textit{joint}
analysis gives intermediate constraints, as it currently uses only the 30 GHz band. At a fixed frequency, however, the \textit{joint} analysis provides stronger constraints than either PS or VID alone, for both astrophysical parameters and the DM-$b$ interaction cross section.

In Fig.~\ref{fig:Bar_Plot}, we show the improvement factor, defined as the ratio of the current strongest $95\%$ C.L. cosmological bounds to the $95\%$ C.L. projected sensitivity obtained from the LIM VID analysis on $\sigma^{\chi p}_{-2}$ and $\sigma^{\chi p}_{-4}$ , for different DM masses. For $n=-2$ and $n=-4$, the strongest existing cosmological bounds considered here are obtained from the Lyman-$\alpha$ forest and CMB+BAO, respectively \cite{Buen-Abad:2021mvc}. An improvement factor greater than unity indicates that the projected LIM sensitivity is stronger than the existing cosmological bound, while a value below unity indicates that the projected sensitivity is weaker than the current bound. For the COMAP-ERA configuration, we find an improvement factor of $4.66$ ($1.78$) at DM mass $m_\chi=1~\mathrm{MeV}$ for $n=-2$ ($n=-4$). At $m_\chi=1~\mathrm{GeV}$, the improvement factor falls below unity for COMAP-EoR. We find that COMAP-ERA  yields better sensitivity than COMAP-EoR, owing to its larger $t_{obs}$. 

\section{LIM Observables}
\label{sec:lim_observables}
We provide a concise overview of the modeling of LIM observables that captures the one-point and two-point statistics of  temperature fluctuations, namely VID and PS, respectively \cite{Sabla:2024dxz}. The temperature fluctuation is $\delta T \equiv T - \langle T \rangle$, where $T$ is the brightness temperature and $\langle T \rangle$ is the mean temperature
\begin{equation}
    \label{eq:mean_temp}
    \langle T(z) \rangle=\frac{c^3(1+z)^2}{8 \pi k_B \nu^3 H(z)}\langle \rho_{\rm L}(z)\rangle= \frac{c^2(1+z)^2}{2k_B\nu^2}\langle I(z) \rangle \,,
\end{equation}
 where $\langle I(z)\rangle=c\langle \rho_{\rm L}(z)\rangle/{4\pi \nu H(z)}$ is the equivalent mean specific intensity, $\nu$ is the rest frame frequency of CO(1-0), $c$ is the speed of light, $z$ is redshift, $k_B$ is the Boltzmann constant, and $H(z)$ is the Hubble parameter. The mean luminosity density is $\langle \rho_{\rm L}(z)\rangle=\int dM L(M,z) \,dn/dM(M,z)$, where $L(M,z)$ is the luminosity as a function of halo mass $M$ and $dn/dM$ is the HMF. These quantities provide the foundation for constructing the PS and VID \cite{Breysse:2022alx,Bernal:2022jap,Bernal:2019jdo}. 

\subsection{Power Spectrum}
\label{subsec:power_spectrum}The primary observable in LIM is the PS, defined as the Fourier transform of the two-point correlation function of $\delta T$, and captures the Gaussian information \footnote{Higher-order correlators \cite{Gil-Marin:2014pva} probe non-Gaussianity.} of intensity maps. The total LIM PS is \cite{Sabla:2024dxz}  \begin{equation}
    \label{eq:tot_LIM_PS}
    P_{\rm tot}(k,\mu,z)= P_{\rm clust}(k,\mu,z)+P_{\rm shot}(z)+P_{\rm noise}\,,
\end{equation}
where $k$ denotes the wavenumber and $\mu = \hat{k}.\hat{k}_{||}$ is the cosine of the angle between the mode and the l.o.s. Since $\delta T $ traces the underlying matter density field, the PS contains a clustering contribution.  The clustering PS is \cite{Bernal:2022jap,Cooray:2002dia,Asgari:2023mej}
\begin{multline}
P_{\rm clust}(k,\mu,z)=
X_{\rm LT}^{2}\int dM\,\frac{dn}{dM}L^{2}\mathcal{U}^{2}(k,M)
\\
+\left[
X_{\rm LT}\int dM\,\frac{dn}{dM}(z)
L\,b_{\rm h}\,F_{\rm RSD}(k,\mu)\,\mathcal{U}(k,M)
\right]^2
P_{\rm m}(k,z),
\label{eq:clust_LIM_PS}
\end{multline}
where the first and second terms describe the one-halo and two-halo terms, respectively. Here, $b_{\rm h}$ is linear halo bias, $\mathcal{U}(k, M)$ is the Fourier transform of the NFW profile of a DM halo of mass M, and $X_{LT}= c^3(1+z)^2/8 \pi k_B \nu^3 H(z)$, a conversion factor between the $T(z)$ and $\rho_{L}(z)$. The brightness temperature maps in LIM are constructed in redshift space, where the l.o.s.\,\,positions of the sources are shifted relative to their real-space positions due to peculiar velocities, giving rise to redshift-space distortions (RSD)~\cite{Bernal:2019jdo}. Their effects are encoded in $F_{\mathrm{RSD}}$~\cite{Hamilton:1997zq,Kaiser:1987qv,Jackson:1971sky}, as explained in detail in the \textit{End Matter}.

The discrete nature of line-emitting sources introduces a scale-independent Poissonian shot-noise term \cite{Breysse_2019,Schaan:2021gzb,Wolz:2017rlw,COMAP:2021rny} in the PS: \begin{equation}
    \label{eq:shot_LIM_PS}
    P_{\rm shot}(z) = X^2_{\rm LT}\int L^2\frac{dn}{dM}dM\,.
\end{equation}
The dependence of both the clustering and shot-noise contributions on cosmological and astrophysical parameters introduces a degeneracy between the two sectors in the LIM PS.
Assuming spatially uniform Gaussian instrumental noise, its contribution to the PS is \cite{Sabla:2024dxz,Bernal:2022jap}
$P_{\rm noise}=V_{\rm vox}\sigma_{\rm N}^{2}$,
where $\sigma_{\rm N}$ denotes the standard deviation of the noise per voxel and $V_{\rm vox}$ is the voxel volume. We discuss this in detail in \textit{Supplemental Material}.

Since $P^{\mathrm{tot}}(k, \mu, z)$ is not directly observable \cite{Bernal:2019jdo} due to the variation of l.o.s.\,\,directions on the sky, we instead consider its monopole\,\cite{Yamamoto:2005dz}, \begin{equation}
    \label{eq:monopole_LIM_PS}
    P_0(k,z)=\frac{1}{2} \int P_{\rm tot}(k,\mu,z) \, d\mu.
\end{equation}
\subsection{VOXEL INTENSITY DISTRIBUTION}
\label{subsec:VID} We outline the framework for computing the one-point probability distribution function (PDF) of the line-intensity field, following halo model formalism and assuming zero-lag clustering~\cite{Breysse:2022alx, Breysse:2016szq, Libanore:2022ntl, Bernal:2023ovz, Sabla:2024dxz}. It is sensitive to the entire line luminosity function, providing a powerful means to break the degeneracy between cosmology and astrophysics. It complements the PS by probing higher-order moments of the luminosity function.

The observable of interest is the fraction of voxels whose measured intensity falls within the $i$-th temperature bin of width $\Delta T_i$, denoted by $B_i$:
\begin{equation}
B_i=\int_{\Delta T_i} dT\,\mathcal{P}_{\rm tot}(T)\,,
\end{equation}
where $\mathcal{P}_{\rm tot}(T)=(\mathcal{P}_{\rm noise}*\mathcal{P})(T)$ is the PDF of the observed brightness temperature, given by the convolution of the signal-only PDF $\mathcal{P}(T)$ with the instrumental-noise PDF $\mathcal{P}_{\rm noise}(T)$, which is assumed to be Gaussian. Unlike standard approaches that assume each emitter contributes only to its host voxel only, we account for spatially extended emission due to finite instrumental resolution~\cite{Sabla:2024dxz, Bernal:2023ovz}. Assuming an intrinsic Dirac delta line profile, the observed intensity distribution is determined by a three-dimensional Gaussian response characterized by the frequency resolution, $\delta\nu$, along the l.o.s.\,\,and the beam full width at half maximum, $\theta_{\rm FWHM}$, in the transverse direction. The profile defines an effective emission volume $V_{\rm prof}$ \cite{Sabla:2024dxz, Bernal:2023ovz}.

The Fourier transform of the brightness-temperature PDF is defined as the characteristic function, with $\tau$ being the Fourier conjugate of the brightness temperature. Given that $\tilde{\mathcal{P}}_{1}^{M}(\tau)$ is the characteristic function from a single halo of mass $M$ and Poisson statistics for the halos, the characteristic function for the full unclustered population, $\tilde{\mathcal{P}}^{u}(\tau)$, becomes~\cite{Bernal:2023ovz}, \begin{equation}
\label{eq:tot_charac_func_unclust}
\tilde{\mathcal{P}}^{(u)}(\tau) = \exp\left[
\int dM\, \frac{dn}{dM} \, V_{\mathrm{prof}} \left( \tilde{\mathcal{P}}^{(M)}_{1}(\tau) - 1 \right)
\right]\,,
\end{equation} 
The halo mass bins are assumed to have infinitesimal widths, such that each mass bin contains either zero or one emitter~\cite{Bernal:2023ovz}. 

Clustering is incorporated as a modulation of the halo abundance, $dn/dM\rightarrow dn/dM(1+\delta_{\rm h}(\boldsymbol{x},M))$ where halo overdensity field $\delta_{\rm h} \equiv (n_{\rm h} - \langle n_{\rm h}\rangle) /\langle n_{\rm h} \rangle = b_{\rm h} \delta_m$ which is a function of position $\boldsymbol{x}$ and halo mass $M$. Here $n_{\rm h}$ is the number density of halos and $\delta_m$ is the matter density fluctuation. Under this prescription, the characteristic function of $\tilde{\mathcal{P}}(\tau)$  for the  $\mathcal{P}_{\rm tot}(T)$ is        \cite{Bernal:2023ovz}, \begin{equation}
\label{eq:tot_charac_func}
\frac{\tilde{\mathcal{P}}(\tau)}{\tilde{\mathcal{P}}^{(u)}(\tau)} =
\exp\left\{
\left[
\int dM \,\frac{dn}{dM} \, V_{\rm prof}
\left(\tilde{\mathcal{P}}^{(M)}_1(\tau) - 1\right) b_{\rm h}
\right]^2
\frac{\sigma^2}{2}
\right\} \,,
\end{equation} where $\sigma^2$ is the zero-lag variance of the matter distribution~\cite{Bernal:2023ovz} smoothed
over the voxel scale in redshift space. 


\section{DM-BARYON INTERACTION}
\label{subsec:DMB}
We investigate non-gravitational elastic scattering between DM and baryons, in a model-independent framework. These interactions permit exchange of heat and momentum between the DM and baryonic fluids. Prior to recombination, baryons are tightly coupled to photons and experience radiation pressure, which can be partially transferred to  DM through DM-$b$ scattering. This inhibits the growth of matter perturbations and thus $P_{\rm m}(k)$, consequently suppressing the LIM observables~\cite{Sabla:2024dxz}.

These effects are incorporated by introducing additional collision terms in the cosmological perturbation equations, following
~\cite{Buen-Abad:2021mvc,Nguyen:2021cnb,Xu:2018efh,Boddy:2018kfv,Nadler:2019zrb,He:2025npy,Boddy:2022tyt,Chen:2002yh,Zentner:2022xux,Li:2022mdj,Rogers:2021byl,DES:2020fxi}.
We focus on interaction models characterized by $\sigma^{\chi b}_{T, n}(v)=\sigma^{\chi b}_{n}v^n$, for the following cases: \textit{i)} \textit{contact}  ($n=0$), \textit{ii)} \textit{electric dipole-moment}-like ($n=-2$), \textit{iii)} \textit{Coulomb}-like ($n=-4$) interactions. 
\begin{figure}[htb]
    \centering
    \includegraphics[width=1.0\linewidth]{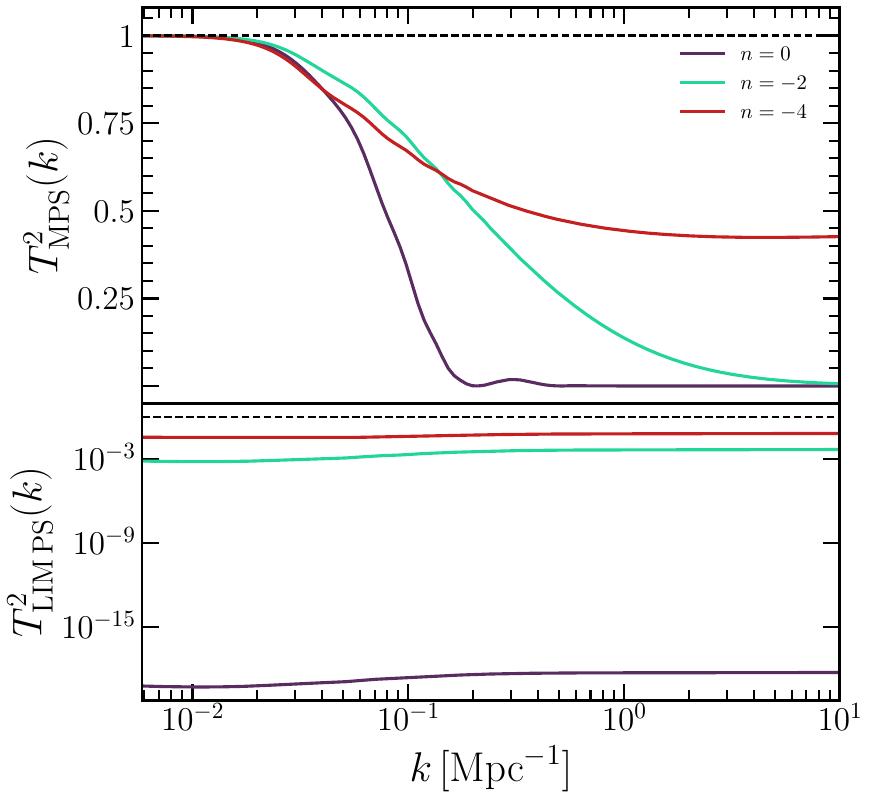}
    \caption{(\textbf{Top}) The ratio $T^{2}_{\rm MPS}(k)=P_{\rm m,\,IDM}(k)/P_{\rm m,\,\Lambda CDM}(k)$, where $P_{\rm m,\,IDM}(k)$ and $P_{\rm m,\,\Lambda CDM}(k)$ are the linear matter power spectrum for the interacting dark matter (IDM) and $\Lambda$CDM model, respectively. (\textbf{Bottom}) The ratio $T^{2}_{\rm LIM\,PS}(k)=P_{\rm 0,\,IDM}(k)/P_{\rm 0,\,\Lambda CDM}(k)$, where $P_{\rm 0,\,IDM}(k)$ and $P_{\rm 0,\,\Lambda CDM}(k)$ are the monopole LIM PS for the IDM and $\Lambda$CDM model, respectively. In this figure $\sigma^{\chi p}_{n}$ for DM-$p$ interactions are fixed at $10^{-24.41}\,\rm cm^2$, $10^{-32.09}\,\rm cm^2$, and $10^{-39.87}\,\rm cm^2$ for $n = 0$ (\textbf{purple}), $n = -2$ (\textbf{cyan}), and $n = -4$ (\textbf{red}), respectively. These values correspond to the $95\%$ C.L. forecasted upper limits derived in PS analysis using COMAP-ERA configuration for the 30 GHz band.
    }
    \label{fig:Transfer_ERA_DMB_10MeV_30GHz}
\end{figure}
To compute the LIM observables in the presence of DM-$b$ interaction, we use the public code \texttt{lim}\footnote{\url{https://github.com/jl-bernal/lim}}. The $P_{\rm m}(k)$ is obtained from a modified version of the Boltzmann solver \texttt{CLASS}\footnote{\url{https://github.com/kboddy/class\_public/tree/dmeff}} \cite{Diego_Blas_2011}, which incorporates the effects of DM-$b$ interaction. In Fig.~\ref{fig:Transfer_ERA_DMB_10MeV_30GHz}, we present the corresponding squared transfer functions, $T_{\rm MPS}^{2}(k)$ for the $P_{\rm m}(k)$ (top panel) and $T_{\rm LIM\,PS}^{2}(k)$ for the LIM PS (bottom panel), with DM--$p$ scattering for $m_{\chi}=10\,\mathrm{MeV}$ and $n=0,-2,-4$. The scattering cross sections $\sigma^{\chi p}_n$ are set to their $95\%$ forecasted upper limits obtained from the PS analysis for the COMAP-ERA setup at 30 GHz. The suppression in $\mathcal{P}(T)$ is shown in Figs.~\ref{fig:LIM_observables_ERA_DMB_10MeV_30GHz} and~\ref{fig:Consist_Check_ERA_10MeV_30GHz_alln_v30}. 

\section{ASTROPHYSICAL MODELS}
\label{sec:astrophysical_models}
In COMAP-EoR and COMAP-ERA, the observed central frequencies of 30 GHz and 16 GHz band correspond to CO(1-0) line from redshifts $z \sim 3$ and $z \sim 6$, respectively. For the 30 GHz band, we model the CO luminosity--halo mass relation using a double power-law parameterization\,\cite{Sabla:2024dxz, COMAP:2021lae} 
\begin{equation}
\label{eq:lumi_30GHz}
\frac{L_{\rm CO} (M)}{L_{\odot}} = 4.9 \times 10^{-5}\times C
\left[
\left(\frac{M}{M_*}\right)^A +
\left(\frac{M}{M_*}\right)^B
\right]^{-1} \,.
\end{equation} 
This form is motivated by semi-empirical galaxy–halo models from UniverseMachine \cite{Behroozi_2019} and constrained by observations of the CO luminosity function from COLDz \cite{Riechers:2018zjg} and the CO PS measured by COPSS  \cite{Keating:2016pka}. Based on these results, we adopt the fiducial parameters: $A=-2.85$, $B=-0.42$, $C=10^{10.63}$, and $M_*=10^{12.3}$ $M_{\odot}$.

For the 16 GHz band, we adopt the CO luminosity–halo mass relation from Yang et al.\,\cite{Yang_2021,Yang_2022}, which is based on semi-analytic galaxy-formation modeling, where
\begin{equation}
\label{eq:lumi_16GHz}
\frac{L_{\rm CO}(M)}{L_{\odot}} =
2N \,\frac{M}{M_{\odot}}
\left[
\left(\frac{M}{M_1}\right)^{-\alpha}
+
\left(\frac{M}{M_1}\right)^{\beta}
\right]^{-1}
\end{equation}
with fiducial parameters, $M_1\simeq10^{11.3}\,M_{\odot}$, $N\simeq10^{-6.8}$, $\alpha\simeq1.7$, and $\beta\simeq0.18$ at $z\sim6$. 

We use luminosity scatter values of $\tilde{\sigma}\simeq0.42$ dex (at 30 GHz) and $\tilde{\sigma}\simeq0.16$ dex (at 16 GHz), following the respective luminosity models, where $\tilde{\sigma}$ quantifies the intrinsic log-normal scatter in CO luminosity at a fixed halo mass, reflecting the stochastic nature of galaxy formation and line-emission processes.
\begin{figure*}
	\begin{center}
		\includegraphics[width=0.99\columnwidth]{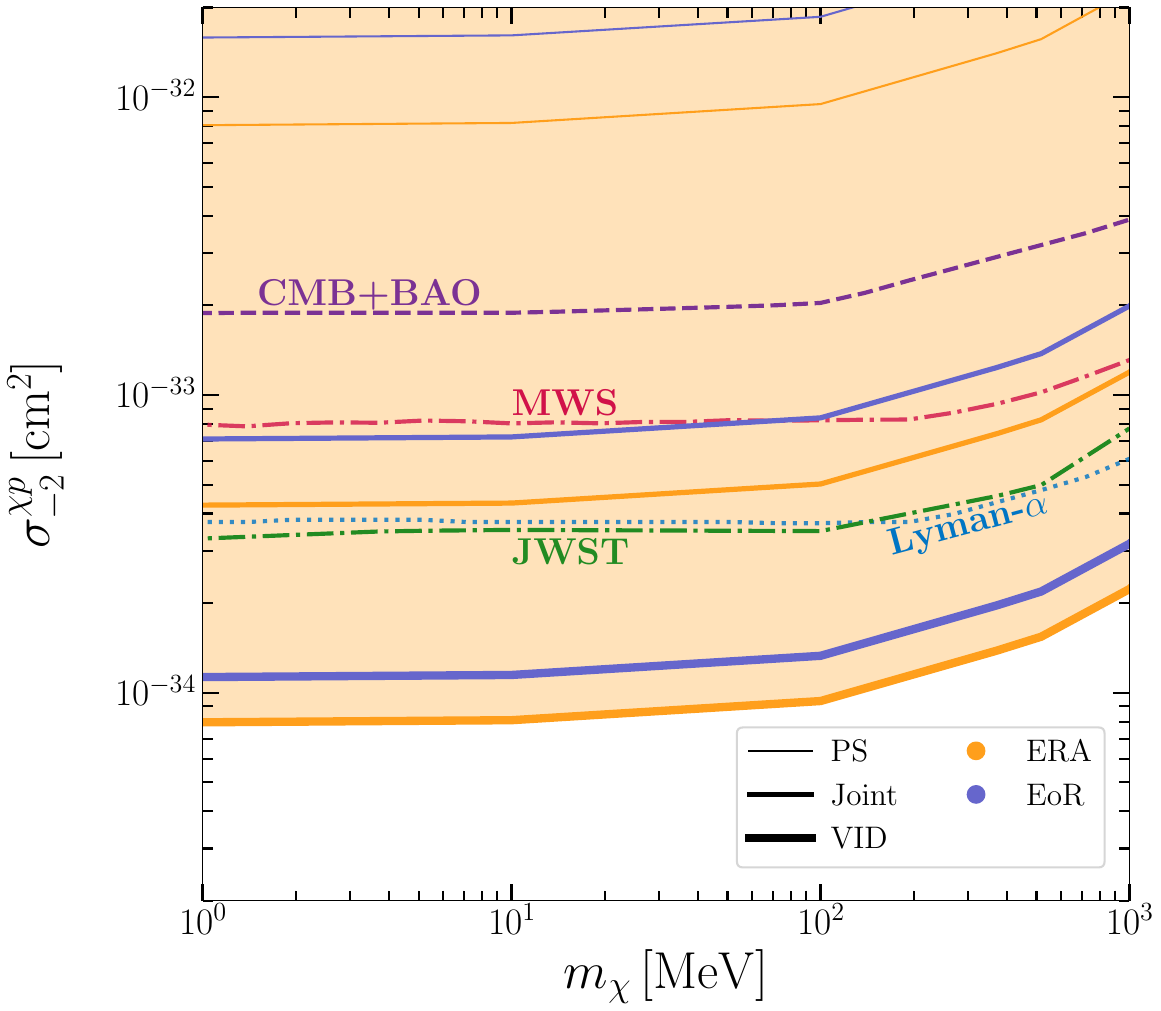}~~
		\includegraphics[width=0.99\columnwidth]{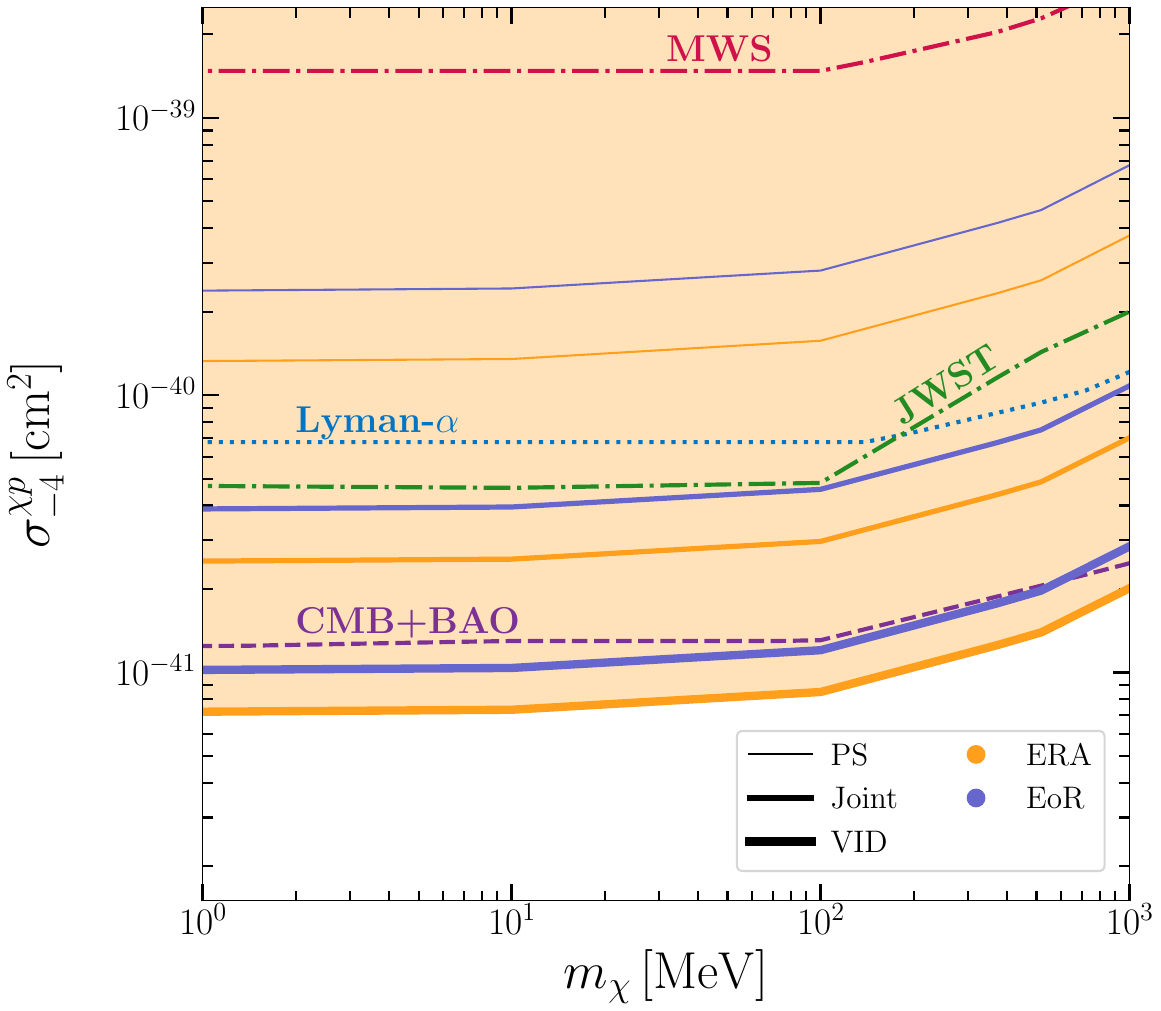}~~\\
		\caption{
		Projected 95\% C.L. upper limits on the normalization of DM-$p$ scattering cross sections
    $\sigma^{\chi p}_{-2}$ (\textbf{left}) and $\sigma^{\chi p}_{-4}$ (\textbf{right}). \textbf{Blue (orange)} curves correspond to upper limits from the COMAP-EoR (COMAP-ERA) survey configuration, while the \textbf{thin}, \textbf{intermediate-thickness}, and \textbf{thick} curves denote forecasts from the \textit{combined} PS, \textit{joint} PS+VID for the 30 GHz band, and \textit{combined} VID analyses, respectively. The orange shaded region is disfavored by our \textit{combined} VID analysis for the COMAP-ERA configuration and gives the strongest projected limits. For comparison, we show existing constraints from other observables. The purple dashed line is the constraint from CMB+BAO, while the dotdashed magenta, dotted blue, and dotdashed green lines show constraints from MWS, Lyman-$\alpha$ forest~\cite{Buen-Abad:2021mvc} and JWST UVLF spectroscopic data, \cite{Das:2025bnr} respectively.
		}
		\label{fig:n-2_n-4_DMB_constraints}
	\end{center}
\end{figure*}
\section{RESULTS}
\label{sec:results} 
We obtain projected $95\%$ C.L.\,\,upper limits on $\sigma^{\chi b}_n$ for COMAP-EoR and COMAP-ERA configurations separately, using the Fisher matrix formalism. In our forecasts, $\sigma^{\chi b}_n$ and the astrophysical parameters are allowed to vary, while the six $\Lambda \rm CDM$ parameters and the sum of neutrino masses are fixed to their best-fit values from the \textit{Planck} 2018 TT, TE, EE+lowE+lensing+BAO
results \cite{Planck:2018vyg}. The astrophysical fiducial values are given in the previous section and we set the fiducial $\sigma^{\chi b}_n$ to zero, corresponding to the $\rm \Lambda CDM$ limit. 

For the PS analysis, we consider the range $ k \in [0.006,10]\,{\rm Mpc}^{-1}$, while for the VID analysis we use $T\in[0,100]\,\mu{\rm K}$. We also consider mean subtraction in the PDF of temperature in our VID analysis. The halo-mass integrals are evaluated over a mass range $M\in[10^{10},10^{15}]\,M_\odot$, with $10^{10}\,M_\odot$ adopted as the minimum mass of appreciable CO-emitting halos \cite{Sabla:2024dxz,Breysse:2021ecm}. We use \textit{top-hat} filter function and \textit{Tinker} \cite{Tinker:2008ff} fitting function for HMF. We adopt \textit{Tinker10} \cite{Tinker_2010} as bias-fitting function for $b_{\rm h}$ \cite{Bernal:2019jdo}. Foreground contamination and line interlopers are assumed to be efficiently mitigated using established techniques \cite{Breysse:2015baa,Sun_2018, Cheng:2020asz, Cunnington:2023jpq, Moriwaki:2020bpr, VanCuyck:2023uli}, and their impact is neglected in our forecasts. We also ignore the effect of line broadening \cite{Bernal:2022jap} due to the peculiar velocity of the gas. We focus on the range $1\,\rm MeV\lesssim \textit{m}_{\chi}\lesssim 1\,\rm GeV$, where cosmological probes—such as the Lyman-$\alpha$ forest, CMB+BAO, MWS, and UVLF measurements are sensitive. We do not extend our analysis below $m_{\chi}\sim 1\rm \,MeV$, as Big Bang nucleosynthesis (BBN) and CMB measurements impose strong bounds in this regime through $\Delta N_{\rm eff}$.
 
 In Fig.~\ref{fig:n-2_n-4_DMB_constraints}, we show $95\%$ C.L. projected upper limits on $\sigma^{\chi p}_{-2}$ (left panel) and $\sigma^{\chi p}_{-4}$ (right panel), for both COMAP-EoR (blue) and COMAP-ERA (orange) surveys. We present our results for $\sigma^{\chi p}_0$ and for DM-$e$ scattering in \textit{Supplemental Material}. For each experiment, we present our results for the \textit{combined} PS (thin solid line), \textit{combined} VID (thick solid line), and \textit{joint} (PS+VID) (intermediate-thick line) analyses for the 30 GHz band. Comparing the sensitivities from the two experiments, we find that COMAP-ERA puts stronger limits on $\sigma^{\chi b}_n$, primarily due to its longer $t_{\rm obs}$.

We find that the \textit{combined} VID analysis substantially improves the constraints compared to the \textit{combined} PS e.g., for the $n=-2$ case, VID tightens the projected limits by two orders of magnitude at $m_{\chi}=1\, \rm MeV$ for both COMAP-ERA and COMAP-EoR. The VID analysis gives stronger projected limits due to its enhanced sensitivity to astrophysical parameters and its ability to capture non-Gaussian information of the intensity field, making it more efficient at breaking degeneracies between astrophysical and cosmological parameters.
In contrast, the PS is more susceptible to degeneracies between astrophysical and cosmological parameters, which limits its constraining power.

For the \textit{joint} analysis in the 30 GHz band, we adopt the covariance matrix provided in the GitHub repository of \texttt{lim} code for COMAP-Y5. Since we use the same window functions, survey-field size, and voxel volume as COMAP-Y5 in the 30 GHz case, the integration entering Eq.~\ref{sigma_Bi_P0} is identical in the two cases, making the COMAP-Y5 covariance directly applicable to our analysis. Since we do not perform the \textit{joint} analysis using the $16+30\,\rm GHz$ bands, the resulting \textit{joint} projected constraints are weaker than the \textit{combined} VID-only sensitivities. In Figs.~\ref{fig:Triangle_Plot_ERA_30GHz_DMB} and \ref{fig:Triangle_Plot_ERA_30GHz_DMe} we see, for a given frequency band including the PS with the VID \cite{Sabla:2024dxz} improves constraints, compared to individual analyses, reflecting the complementarity of the two statistics. Their different sensitivities to astrophysical and cosmological effects allow the \textit{joint} analysis to further reduce parameter degeneracies and tighten the sensitivities. We therefore expect that a \textit{combined} \textit{joint} analysis incorporating both frequency bands would further improve our results. In Fig.~\ref{fig:n-2_n-4_DMB_constraints}, we see that our projected limits on $\sigma^{\chi p}_{-2}$ and $\sigma^{\chi p}_{-4}$ cases, for VID-only analysis, is stronger than the existing limits for both COMAP-EoR and COMAP-ERA. The relative constraining power of these various probes is determined by the range of scales they access, their associated statistical and systematic uncertainties, and the extent to which the corresponding observables respond to DM-$b$ interactions relative to the $\Lambda$CDM prediction.

\section{CONCLUSIONS}
\label{sec:conclusions}
For the \textit{first time}, we forecast the sensitivities of next-generation CO experiments, COMAP-EoR and COMAP-ERA, to DM–b interactions.\,\,For a broad class of particle physics models, the DM–b momentum-transfer cross-section can be parametrized as $\sigma^{\chi b}_{T,n}=\sigma_n^{\chi b}v^n$, with $n=0,-2,$ and $-4$. Such interactions suppress the $P_{\rm m}(k)$ on small scales, which is subsequently reflected in the reduced abundance of low-mass halos hosting CO-emitting galaxies. Our analysis considers both the PS and VID, which carry distinct information about the underlying cosmology and line-emission astrophysics. We find that the resulting forecasts can significantly improve upon existing cosmological bounds, particularly for \textit{electric dipole-moment}-like $n=-2$ and \textit{Coulomb}-like $n=-4$ interactions between DM and protons for sub-GeV DM masses. This demonstrates the potential of LIM to provide competitive constraints on non-gravitational interactions of DM.

Alongside its many advantages, LIM comes with its own set of caveats, casting subtle shadows over its promise as a cosmological probe. One of the most important is disentangling foregrounds from the cosmological signal~\cite{Chang:2026ake, Bernal:2022jap, Schaan:2021hhy}. Residual foreground contamination can act as an additional source of noise, increasing the uncertainties of LIM observables beyond those arising from instrumental noise and sample variance alone, and thereby degrading sensitivity to the underlying cosmological signal~\cite{Chang:2026ake, Schaan:2021hhy}. Nevertheless, the spectral smoothness of many continuum foregrounds can be exploited to model and reduce their contribution~\cite{Schaan:2021hhy, Chang:2026ake, McQuinn:2005hk, Jelic:2008jg, Jelic:2008jg}. We discuss some of the potential foreground contaminations in \textit{End Matter}.

The rich astrophysics encoded in line emission also provides opportunities to further improve the cosmological information extracted from LIM~\cite{Bernal:2023ovz, 1985ApJ...291..722T, Stacey:2018yqe,  Dame:2000sp, Walter:2003zh, 2011MNRAS.412.1913I, 2011MNRAS.415...32S}. Different emission lines probe distinct phases and physical conditions of the interstellar and intergalactic media, providing information on electron density, metallicity~\cite{2012ApJ...746...69G, Popping:2013ixa, 2023ApJ...942...24S}, ionization state, and gas pressure~\cite{Bernal:2022jap, Chang:2026ake}. Since these lines contain complementary astrophysical information while tracing the same underlying matter distribution, their cross-correlations provide a promising route to disentangle astrophysical uncertainties from cosmological effects, which is particularly relevant for strengthening constraints on DM–b interactions. Moreover, cross-correlating lines from the same cosmological volume can suppress interloper contributions, providing additional robustness against contaminants~\cite{Chang:2026ake}.

Another avenue for improving robustness against foregrounds and systematics is to cross-correlate LIM with external tracers, including galaxy and quasar catalogs~\cite{Alonso:2017dgh, Modi:2021okf, 2019MNRAS.490..260B}, the cosmic infrared background, weak lensing~\cite{Foreman:2018gnv}, and CMB secondary anisotropies such as  kinetic-Sunyaev–Zel’dovich-effect (kSZ), CMB lensing~\cite{Schaan:2021gzb}.\,\,In particular, CMB lensing and weak lensing trace the projected matter distribution more directly than galaxy tracers, reducing their dependence on galaxy bias and uncertainties in the galaxy–halo connection.~\cite{Bernal:2023ovz, Chung:2022lpr, Singh:2016xey}. Combined with the redshift information from LIM, these cross-correlations can provide a tomographic view of the matter distribution and help disentangle cosmology from line-emission astrophysics~\cite{Bernal:2023ovz, Chung:2022lpr, Chang:2026ake}.

The projected limits derived in this work can also be used to constrain particle physics models of DM. Following Ref.~\cite{Chu:2018qrm, Buen-Abad:2021mvc}, we consider a model in which Dirac fermion DM interacts electromagnetically through an electric dipole moment (EDM), and compare the projected sensitivity to the DM EDM with existing constraints in the \textit{Supplemental Material}. This interaction leads to an $n=-2$ velocity dependence of the momentum-transfer cross section, one of the cases for which LIM provides our the strongest projected sensitivity. Although LIM also provides strongest projected constraints for the $n=-4$ case compared with existing limits, we do not map these constraints onto a specific particle physics model because of the nontrivial effects coming from forward scattering and Debye screening in \textit{Coulomb-like} interactions~\cite{Buen-Abad:2021mvc}.

With many ongoing and upcoming mega-surveys, this is an exciting time for cosmology and astrophysics. Beyond advancing our understanding of these fields, these surveys can open new directions for DM searches, complementary to conventional direct and indirect detection. Combining probes with complementary observables and largely uncorrelated systematics can further provide independent handles on physics beyond the Standard Model~\cite{Bernal:2023ovz, Bernal:2022wsu}. Beyond DM–$b$ interactions, LIM can probe a broader landscape of fundamental physics, including DM decay and annihilation~\cite{Creque-Sarbinowski:2018ebl, Bernal:2020lkd, Shirasaki:2021yrp}, dark energy~\cite{MoradinezhadDizgah:2023src}, neutrinos~\cite{Shmueli:2024npx}, and primordial non-Gaussianity~\cite{Sabla:2024dxz}. We look forward to this rapidly evolving landscape, where the growing potential of intensity mapping may open new avenues for deepening our understanding of the Universe.

\section{ACKNOWLEDGMENTS}

We are especially grateful to Vivian Sabla for her insightful responses to our queries, which significantly guided the direction of this project. We thank Priyank Parashari for some very useful discussions. We thank Aditya Srivatsa, Akash Kumar Saha, Anirban Chakraborty, Anirban Das, Debajit Bose, Deep Ghosh, Deep Jyoti Das, Jaya Doliya, K Parthsarthi, Manas Mohit Dosibhatla, Mayuri S. Rao, Rinchen Sherpa, Sanskrita Chakraborty, Saurabh Singh, Shadab Alam, Shamim Akhtar, Shiriny Akhtar, Souradeep Das, Subhadip Bouri, Suman Majumder, Tanoy Sarkar, Tarak Nath Maity, Tarun Souradeep and Yashrajsinh Mahida for discussions. We also acknowledge the use of ChatGPT, Claude, and Perplexity as AI-assisted tools during the preparation of this manuscript, including for language editing, brainstorming, literature searches, and assistance with code development and debugging. All scientific content, analysis, and conclusions were independently verified by the authors. D.G. acknowledges
the financial support provided by the Ministry of Education (MoE), Government of India. R.M.\,\, is supported by the ISRO-IISc STC Grant No. ISTC/PHY/RL/499 and ANRF National Postdoctoral Fellowship (NPDF) under Project No. PDF/2025/001161.  R.L. acknowledges financial support
from the institute start-up funds and ANRF for the grant
no. ANRF/ARG/2025/005140/PS.  We thank Nirmal Raj for allowing us to use his workstations for some of the computations performed in this work. We would like to thank Vinayak Sanil, who maintained the computing workstations, which was invaluable for various computations performed in this work. 

\bibliographystyle{JHEP}
\bibliography{ref} 

\clearpage

\section{End Matter}

\renewcommand{\theequation}{A\arabic{equation}} 
\setcounter{equation}{0} 
\renewcommand{\thetable}{A\arabic{table}} 
\setcounter{table}{0} 
\subsection{Fisher Forecast}
\label{sec:app_Fisher_Forecast}
Fisher matrix techniques are used widely in astrophysics and cosmology \cite{dafebb01-f6c3-3bb0-8577-2b844df40cce, Ade_2019, coe2009fishermatricesconfidenceellipses} to forecast the precision of future experiments while they are
still in the design phase. The Fisher approach quantifies the minimum uncertainty of parameters from given observations (mock) of a future experiment. It examines the curvature of the likelihood function in the parameter space around its maximum. A steeper curvature corresponds to tighter constraints. As such, the Fisher formalism offers a convenient framework for assessing the constraining power of planned observations and for guiding experimental design. Formally, the Fisher matrix element is defined by,
\begin{equation}
F_{\rm ij} = - \left\langle \frac{\partial^2 \ln \mathcal{L}}{\partial \theta_{\rm i} \, \partial \theta_{\rm j}} \right\rangle\,,
\label{eq:fisher_def}
\end{equation}
where $F_{\rm ij}$ denotes the 
(i, j) element of the Fisher matrix, $\theta_{\rm i}$ represents the model parameters, $\mathcal{L}$ is the likelihood function, and the angle brackets indicate an ensemble average over different realizations of the data.
Assuming a Gaussian likelihood and a covariance matrix of data $\zeta$, one can show that, with some approximation, the Fisher information matrix takes the following form \cite{Sabla:2024dxz}: \begin{equation}
    F_{\rm ij} = \frac{\partial \Theta^T}{\partial \theta_{\rm i}}\zeta^{-1}\frac{\partial \Theta}{\partial \theta_{\rm j}}\,,
\end{equation}
where $\Theta$ is our observables
and $\theta$ denotes model parameters. For the analysis where observed frequency is 30 GHz, $\theta_{\rm i}=\{A,B,{\rm log}\,C,{\rm log}\,M_*,\sigma^{\chi b}_n\}$ 
and for the 16 GHz observed frequency case $\theta_{\rm i} = \{{\rm log}\,M_1,{\rm log}\,N, \alpha, \beta, \sigma^{\chi b}_n\}$.
For a combined analysis of the PS signal from both frequency bands in COMAP-EoR and COMAP-ERA configuration, the Fisher Matrix takes the form
\begin{equation}
    F_{\rm ij} = \frac{\partial \tilde{P}^T_0}{\partial \theta_{\rm i}}\zeta^{-1}_{\tilde{P}_0} \frac{\partial \tilde{P}_0}{\partial\theta_{\rm j}}\,,
    \label{eq:FF_P0}
\end{equation}
where $\theta_{\rm i}=\{A,B,{\rm log}\,C,{\rm log}\,M_*, {\rm log}\,M_1,{\rm log}\,N, \alpha, \beta,\sigma^{\chi b}_n\}$, and $\tilde{P}_0$ denotes the window function convolved (for further details, see the \textit{Supplemental Material}) monopole PS vector, $[\tilde{P}_{0,30}, \tilde{P}_{0,16}]$. The covariance matrix is given by, \begin{equation}
    \zeta_{\tilde{P}_0} = \begin{bmatrix}
        \mathcal{C}_{\tilde{P}_{0,30}}&0 \\
        0&\mathcal{C}_{\tilde{P}_{0,16}}
    \end{bmatrix}\,,
    \label{eq:cov_PS_30_16}
\end{equation}
 where $\mathcal{C}_{\tilde{P}_{0,30}}$, $\mathcal{C}_{\tilde{P}_{0,16}}$ denote the covariance (for further details, see the next section) blocks of the monopole PS at 30 GHz and 16 GHz, respectively. Similarly, for the combined VID analysis the Fisher Matrix will take the form \begin{equation}
    F_{\rm ij} = \frac{\partial B^T_i}{\partial \theta_{\rm i}}\zeta^{-1}_{B_i} \frac{\partial B_i}{\partial\theta_{\rm j}}\,,
    \label{eq:FF_P0}
\end{equation} and the covariance matrix will be
\begin{equation}
    \zeta_{B_i} =
    \begin{bmatrix}
        \mathcal{C}_{B_{i,30}} & 0 \\
        0 & \mathcal{C}_{B_{i,16}}
    \end{bmatrix}\,,
    \label{eq:cov_VID_30_16}
\end{equation}
where $\mathcal{C}_{B_{i,30}}$ and $\mathcal{C}_{B_{i,16}}$ denote covariance blocks  of the VID bin at 30 GHz and 16 GHz, respectively.  We consider 250 wavenumber bins for the PS and 250 temperature bins for the VID. In this analysis, both covariance matrices are taken to be diagonal, neglecting correlations between the two frequency bands as well as mixing between different temperature bins in the VID and different $k$ bins in the PS for simplicity. The vanishing off-diagonal blocks in Eqs.~\eqref{eq:cov_PS_30_16},\eqref{eq:cov_VID_30_16} indicate that the two frequency bands are assumed to be independent.

For the \textit{joint} analysis of the data from the telescope where obserevd frequency is 30 GHz, we construct the observable vector $\Theta = [\tilde{P}_{0,30},B_{i,30}]$ \cite{Sabla:2024dxz}. The corresponding covariance matrix is \begin{equation}
    \zeta_{\rm \textit{Joint}} = \begin{bmatrix}
        \mathcal{C}_{\tilde{P}_{0,30}}&\mathcal{C}_{\tilde{P}_{0,30},B_{i,30}} \\
        \mathcal{C}_{B_{i,30},\tilde{P}_{0,30}}&\mathcal{C}_{B_{i,30}}
    \end{bmatrix}\,.
    \label{eq:cov_PS_Bi_30}
\end{equation}
Here, $\mathcal{C}_{\tilde{P}_{0,30}}$ and  $\mathcal{C}_{B_{i,30}}$ are the covariance block of $\tilde{P}_0$ and $B_i$ at 30 GHz respectively and the off-diagonal blocks represent the covariance between $\tilde{P}_0$ and $B_i$ at 30 GHz. 

The Fisher matrix provides estimates of parameter uncertainties as well as their covariances. The covariance matrix of parameters is given by the inverse Fisher matrix, $\tilde{\zeta}_{\rm ij}=(F^{-1})_{\rm ij}$. A useful measure of parameter degeneracy is the correlation coefficient, defined as
\begin{equation}
\rho_{\rm ij} = \frac{\tilde{\zeta}_{\rm ij}}{\sqrt{\tilde{\zeta}_{\rm ii} \tilde{\zeta}_{\rm jj}}}\,,
\end{equation}
which lies in the range $[-1,1]$. Positive (negative) values correspond to positively (negatively) correlated parameters, while $\rho_{\rm ij}=0$ indicates no correlation.
The correlation coefficient quantifies degeneracies between parameters: Large values of $|\rho_{\rm ij}|$ correspond to strong degeneracies, where changes in one parameter can be compensated by changes in another, leading to similar fits to the data and making their independent determination more challenging.
\subsection{Covariance}
\label{sec:app_Cov}
The covariance of the LIM PS receives contributions from sample variance, instrumental noise, and residual contamination from foregrounds and line interlopers \cite{Oxholm:2022mnn} that remain after the cleaning procedure. Foreground contamination is assumed to be effectively controlled using conventional techniques \cite{Bernal:2019jdo}. A detailed discussion of foreground and line-interloper mitigation techniques are discussed in \cite{Lidz:2016lub, Oh:2003jy, Wang:2005zj, Liu_2011, Cheng:2016yvu, Sun_2018, Breysse:2015baa}, and are beyond the scope of this work. Neglecting coupling between Fourier modes~\cite{1994ApJ...426...23F}, the monopole PS variance per $k$ bin is given by \cite{Sabla:2024dxz, Bernal:2022jap},
\begin{equation}
\sigma^2_{\tilde{P}_0}(k, z) = \frac{1}{2N_{\mathrm{modes}}}
\int_{-1}^{1} \left[ \tilde{P}_{\rm tot}(k,\mu,z)\right]^2 d\mu \,,
\label{eq:ps_variance}
\end{equation}
where the number of independent modes in the survey volume $V_{\rm field}$ within a bin of width $\Delta k$ is $N_{\rm modes} = (k^2 \Delta k/4 \pi^2)V_{\rm field} N_{\rm field}$ with $N_{\rm field}$ the total number of fields. 
In our analysis, we have a survey area of $4\,\rm deg^2$ per field and three independent fields for both the experiments.

For the VID covariance, we adopt the diagonal variance \cite{Sabla:2024dxz} \begin{equation}
    \sigma^2_{B_i} = \frac{B_i(1-B_i)}{N_{\rm vox}}\,,
\end{equation} 
where $N_{\rm vox}$ is the total number of voxels across all three independent survey fields. This expression corresponds to the statistical counting fluctuations associated with the finite number of voxels. Numerical simulations of spatially correlated lognormal realizations show that the multinomial approximation, which treats the numbers of voxels falling into different temperature bins as multinomially distributed counts, provides a good description of the diagonal VID covariance, while the off-diagonal covariance deviates from the multinomial prediction but remains small~\cite{Bernal:2023ovz}.\,\,We therefore neglect correlations between different temperature bins in the VID, following \cite{Sabla:2024dxz}. We also neglect the supersample covariance contribution described in \cite{Sato-Polito:2022fkd}, as it is found to be subdominant to the statistical covariance \cite{Bernal:2023ovz}. 

The covariance between the PS and the VID arises from their common dependence on the underlying density field and the luminosity distribution of line-emitting sources. The VID is sensitive to fluctuations in the number density of emitters,
$\delta_{\rm h}$, while the PS depends on two powers of the temperature
fluctuation, $\delta T$. Consequently, their cross-covariance depends on the bispectrum
$\langle\delta_{\rm h}\delta T\delta T\rangle$. Following
\cite{Sato-Polito:2022fkd, Sabla:2024dxz}, we write the PS--VID
cross-covariance as
\begin{equation}
\begin{aligned}
& \sigma_{B_i \tilde{P}_0} = \frac{N_{\rm vox}}{V_{\rm field}^2} \Upsilon_i
\int \frac{d^2\Omega_{\hat{k}}}{4\pi}
\int \frac{d^3 q_1}{(2\pi)^3}
\int \frac{d^3 q_2}{(2\pi)^3}
\int \frac{d^3 q_3}{(2\pi)^3} \\
&\quad\times
W_{\rm vol}(\mathbf{q}_1)
W_{\rm vol}(\mathbf{q}_2)
W_{\rm vol}(\mathbf{q}_3)
W_{\rm res}(-\mathbf{q}_1)
W_{\rm res}(\mathbf{k}-\mathbf{q}_2) \\
&\quad\times
W_{\rm res}(-\mathbf{k}-\mathbf{q}_3)
\left\langle
\delta_{\rm h}(-\mathbf{q}_1)
\delta T(\mathbf{k}-\mathbf{q}_2)
\delta T(-\mathbf{k}-\mathbf{q}_3)
\right\rangle ,
\end{aligned}
\label{sigma_Bi_P0}
\end{equation}
where $\sigma_{B_i\tilde{P}_0}$ denotes the covariance between the VID bin
$B_i$ and the smoothed monopole PS $\tilde{P}_{0}$ in the wavenumber bin $k$. Here, The window functions $W_{\rm vol}$ and $W_{\rm res}$ account for the effects of the finite survey volume and instrumental resolution, respectively. We discuss this
in detail in \textit{Supplemental Material}.

The factor $\Upsilon_i$ captures the response of the VID in temperature bin $i$ to fluctuations in the local halo abundance, with its magnitude determined by the line-luminosity distribution.
It is given by
\begin{equation}
    \Upsilon_i =
    \int_{\Delta T_{i}} dT
    \int \frac{d\tau}{2\pi}
    \tilde{\mathcal{P}}_{\rm noise}
    \tilde{\mathcal{P}}^{(u)}
    \log\!\left(\tilde{\mathcal{P}}^{(u)}\right)
    e^{iT\tau}\,,
\end{equation}
where $\tilde{\mathcal{P}}_{\rm noise}$ is the characteristic function of the probability distribution for the instrumental noise contribution to the temperature of a voxel. The sign of $\Upsilon_i$ indicates whether the corresponding temperature-bin probability increases or decreases in response to a local halo overdensity or underdensity \cite{Sato-Polito:2022fkd}. For the \textit{joint} PS+VID analysis, $N_{\rm vox}$ and $V_{\rm field}$ are taken to be the same as those used in the corresponding PS and VID analyses. 
\subsection{Redshift Space Distortion}
\label{RSD_Factor}
Redshift-space distortions arise from the fact that observed redshifts receive contributions not only from the Hubble expansion but also from the peculiar velocities of line-emitting halos along the l.o.s. This mapping from real space to redshift space introduces anisotropies in the observed clustering pattern, rendering the PS dependent on $\mu=\hat{k}\cdot\hat{k}_{\parallel}$, where
$\hat{k}_{\parallel}$ denotes the l.o.s.\,\,direction. In eq.\,\eqref{eq:clust_LIM_PS}, this effect is incorporated through the factor $F_{\rm RSD}(k,\mu)$, given by \cite{Bernal:2022jap},
\begin{equation}
\label{rsd_factor}
   F_{\rm RSD}(k,\mu)=\frac{1+\frac{f \mu^2}{b_{\rm h}(M)}}{1+0.5(k\mu\sigma_{\rm pv}(M))^2}\,,
\end{equation}
where $M$ is the mass of a halo, $f$ is the linear growth rate, and $\sigma_{\rm pv}$ is the halo pairwise velocity dispersion. The numerator describes the large-scale Kaiser enhancement \cite{Kaiser:1987qv}, arising from coherent infall toward overdensities, while the denominator models the small-scale suppression due to random virial motions (the Finger-of-God effect) \cite{Jackson:1971sky}. We set $\sigma_{\rm pv}=0$ as in \cite{Sabla:2024dxz}, for simplicity. 
\subsection{Foreground}
Foreground emission is a major systematic challenge for LIM, as emission from astrophysical and terrestrial sources
can substantially exceed the target line signal \cite{Oxholm:2022mnn,Bernal:2022jap}. Continuum foregrounds include zodiacal emission, Galactic dust and
synchrotron radiation, free-free emission, and the CMB, while observations at millimeter wavelengths can also be
affected by atmospheric emission, ground pickup, and radio-frequency
interference. Unlike the LIM signal, which traces spatial fluctuations
in the large-scale matter distribution and therefore varies rapidly
with frequency, continuum foregrounds are generally spectrally smooth. This
spectral coherence provides a means of mitigation by filtering out low-$k_\parallel$ modes, although such procedures can also remove the signal and are sensitive to the frequency
dependence of the instrumental response and beam. In addition, cross-correlation with galaxy surveys can further suppress foreground and instrumental noise contributions, since these contaminants are uncorrelated with the underlying matter density field traced by both surveys.
\clearpage
\onecolumngrid
\begin{center}
\textbf{\large \mytitle}

\vspace{0.08in}
{ \it \large Supplemental Material}\\ 
\vspace{0.12in}

{Durba Ghosh, Ranjini Mondol, and Ranjan Laha}
\end{center}
\twocolumngrid
 
\setcounter{figure}{0}
\setcounter{section}{0}
 
\makeatletter
 
\renewcommand{\theequation}{S\arabic{equation}}
\renewcommand{\thefigure}{S\arabic{figure}}
 
\renewcommand{\theHfigure}{suppfigure.\arabic{figure}}
 
\makeatother
In this work, we investigate LIM as a probe of
DM-$b$ interactions \textit{for the first time} using the upcoming COMAP-EoR and COMAP-ERA surveys. We forecast constraints on the normalization cross-section of the momentum-transfer cross section, $\sigma^{\chi b}_{T,n}=\sigma^{\chi b}_n v^n$ for $b\in(p,e)$ and $n=0,-2,-4$, using the LIM observables PS and VID. We obtain our results using separate PS and VID analyses, each combining the signals from the $30$ and $16\,\mathrm{GHz}$ bands, as well as a \textit{joint} PS+VID analysis using the $30\,\mathrm{GHz}$ signal. We obtain the strongest constraints for DM–p interactions with \textit{electric dipole-moment}-like ($n=-2$) and \textit{Coulomb}-like ($n=-4$) interactions. In the main text, we present the projected sensitivities for these interactions for sub-GeV DM masses and quantify the improvement that LIM can provide over existing limits.

In this \textit{Supplemental Material}, we provide additional details of the theoretical modeling and survey implementation underlying these forecasts. We first discuss the impact of DM-$b$ interactions on the LIM observables, tracing the suppression of the $P_{\rm m}(k)$ and HMF into the intensity statistics. We then detail the experimental specifications adopted for
COMAP-EoR and COMAP-ERA, including the treatment of instrumental resolution, survey geometry, and noise. Finally, we present additional projected constraints for the velocity-independent and velocity-dependent interaction models considered in this work, including DM-$p$ and DM-$e$ scattering, together with marginalized posterior distributions and
joint confidence regions on the interaction and astrophysical parameters. 
 \section{Effects of DM-$b$ on LIM Observables} 
\begin{figure}
    \centering
    \includegraphics[width=1.0\linewidth]{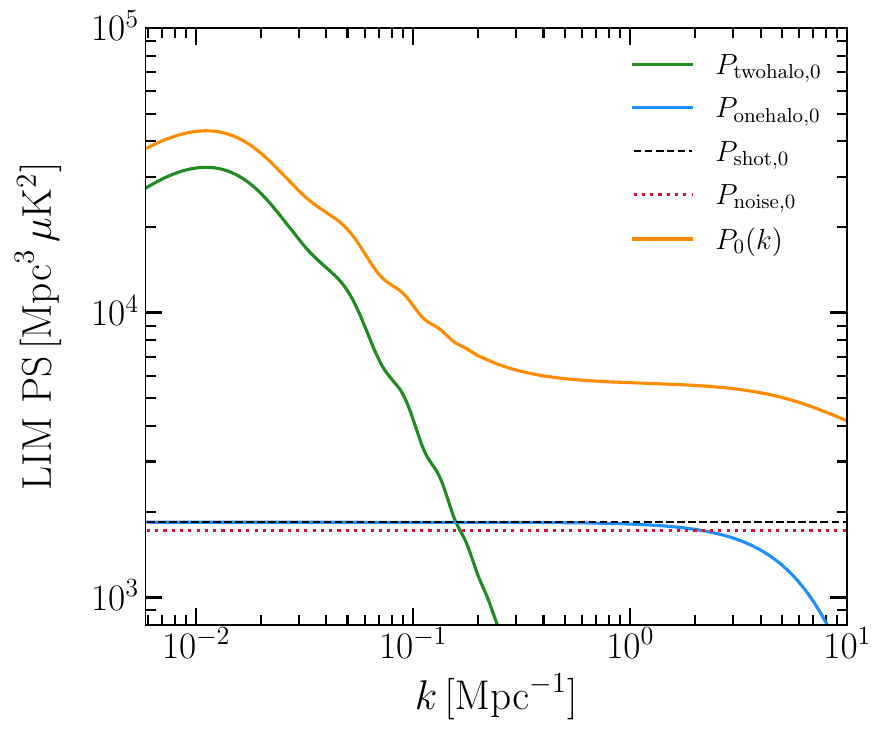}
    \caption{Contributions to the total LIM monopole PS, $P_0(k)$, in the $\Lambda$CDM model for the COMAP-ERA 30 GHz survey. The subscript $0$ denotes the monopole, obtained by angle averaging each component. Shown are the two-halo clustering term (\textbf{green}), one-halo term (\textbf{blue}), shot-noise contribution (\textbf{black dashed}), instrumental noise contribution (\textbf{red dotted}), and the total monopole PS (\textbf{orange}). On large scales ($k\lesssim0.1,{\rm Mpc}^{-1}$), the signal is dominated by the two-halo clustering component, reflecting correlations between distinct halos. At smaller scales, the one-halo and shot-noise contributions become increasingly important, with the shot-noise contribution dominating around the highest $k$ shown.}
    \label{fig:LIM_PS_ERA_30GHz}
\end{figure}

We calculate, for the first time, LIM observables considering DM-$b$ interaction using \texttt{lim} code and modified \texttt{CLASS} code. In Fig.~\ref{fig:LIM_PS_ERA_30GHz} we show the different contributions to the total LIM monopole PS, obtained by angle averaging  the corresponding components of Eq.~\eqref{eq:tot_LIM_PS}, for the case of $\Lambda \rm CDM$ in the COMAP-ERA configuration for 30 GHz band. It is evident that at large scales (low $k$) the two-halo clustering PS dominates over the one-halo and shot noise contributions. At higher $k$ (smaller scales) values, both the shot noise PS, instrumental noise PS and one-halo clustering PS dominate. 

\begin{figure*}
	\begin{center}
		\includegraphics[width=0.99\columnwidth]{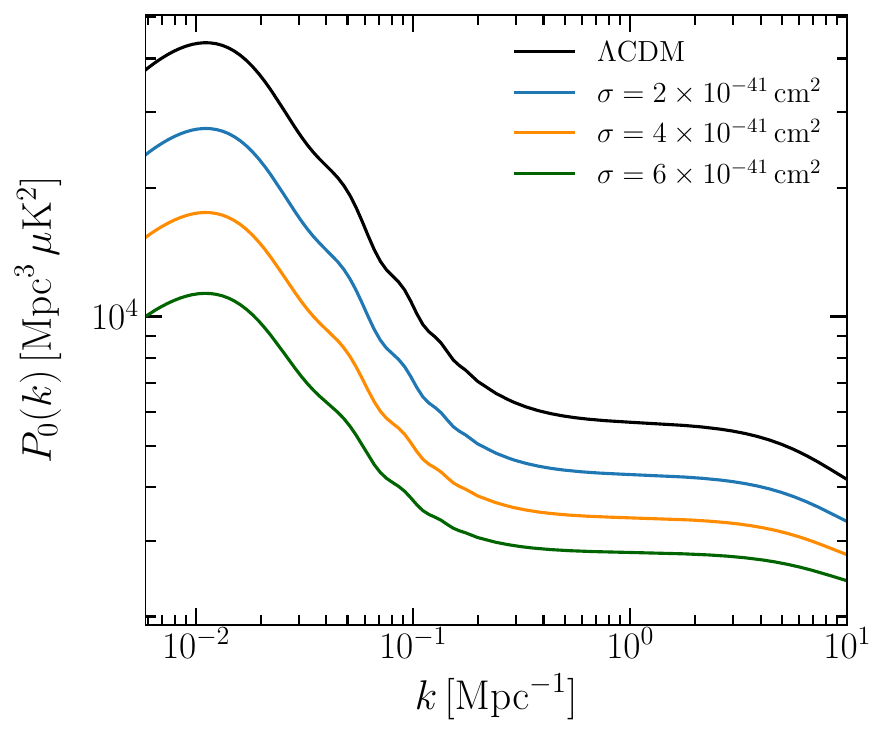}~~
		\includegraphics[width=0.99\columnwidth]{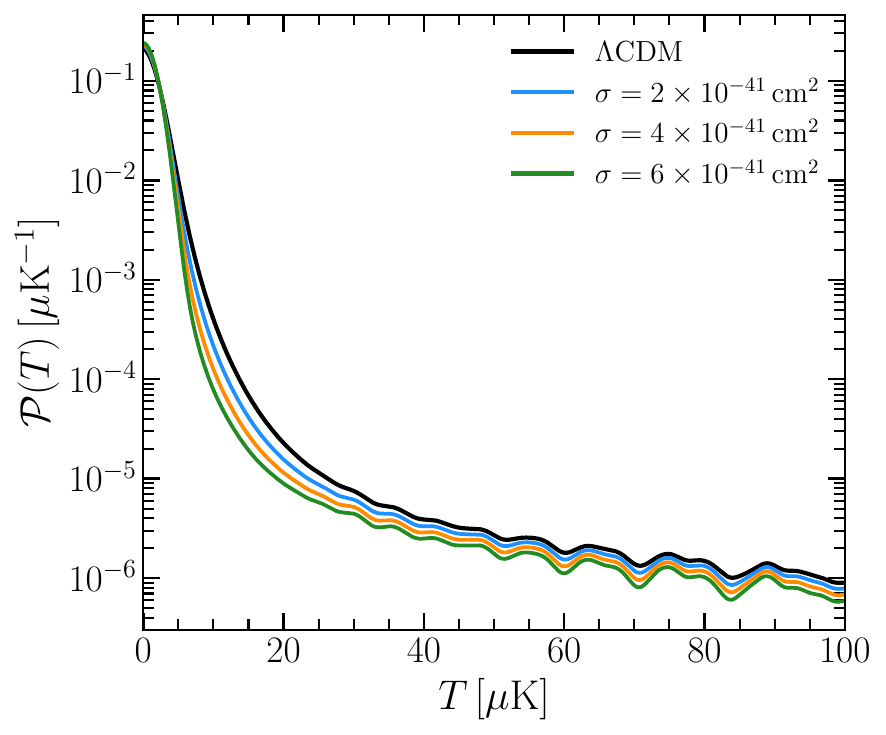}~~\\
		\caption{
		LIM observables for different values of $\sigma^{\chi p}_{-4}$ in the COMAP-ERA 30 GHz configuration, assuming $m_\chi=10\,\rm MeV$. \textbf{\textit{Left:}} The LIM monopole PS $P_0(k)$. \textbf{\textit{Right:}} The temperature PDF $\mathcal{P}(T)$. The \textbf{black} curves correspond to the $\Lambda\rm CDM$ prediction, while the colored curves correspond to DM-$b$ interactions of different strengths. The interactions suppress the LIM PS (left panel) and redistribute the voxel temperatures, with the temperature probability distribution $\mathcal{P}(T)$ shifting toward lower values (right panel). Stronger interactions produce increasingly stronger suppression of $P_0(k)$ and larger deviations of $\mathcal{P}(T)$ from the $\Lambda$CDM prediction.
		}
		\label{fig:LIM_observables_ERA_DMB_10MeV_30GHz}
	\end{center}
\end{figure*}
Fig.~\ref{fig:LIM_observables_ERA_DMB_10MeV_30GHz} shows the total LIM monopole PS $P_0(k)$ (left panel) and temperature probability distribution $\mathcal{P}(T)$ (right panel) for different values of $\sigma^{\chi p}_{-4}$ in the COMAP-ERA 30 GHz configuration for $m_{\chi}=10\,\mathrm{MeV}$. Relative to $\Lambda$CDM, DM-$b$ interactions lead to an overall suppression of the LIM PS, due to the decrease in the number density of halos hosting line-emitting galaxies, thereby suppressing the amplitude and clustering of the corresponding intensity fluctuations.  The suppression is more pronounced on large scales where the two-halo contribution dominates, as also evident from Fig.~\ref{fig:Transfer_ERA_DMB_10MeV_30GHz}.
From Eq.~\eqref{eq:clust_LIM_PS} we see the two-halo PS contains the $P_{\rm m}(k)$ multiplied by a factor involving the HMF and $\mathcal{U}(k, M)$. Since the HMF is modified by DM-$b$ interactions, this multiplicative factor, differs between the $\Lambda$CDM and interacting cases. At small $k$, where $\mathcal{U}(k,M)\simeq1$, this factor is nearly scale independent, and the scale dependence of the interaction-induced suppression in the LIM PS therefore primarily follows that of the suppressed $P_{\rm m}(k)$. However, the two-halo term becomes increasingly subdominant at large scales, while the one-halo, shot-noise, and instrumental-noise contributions become increasingly important. At larger $k$, the scale dependence of $\mathcal{U}(k,M)$ becomes important and $\mathcal{U}(k,M)$ in the one-halo term determines the scale dependence of the total LIM PS. The one-halo and shot-noise
terms are also modified by the interaction through the HMF, but their dependence on the second moment of the luminosity function, which preferentially weights bright and massive halos (Eqs.~\eqref{eq:clust_LIM_PS}, \eqref{eq:shot_LIM_PS}), makes these contributions less sensitive to suppression concentrated in lower-mass halos \cite{Sabla:2024dxz}. Consequently, their fractional changes due to interaction  are smaller than that of the two-halo contribution. This dilutes the interaction-induced suppression at high $k$. The total LIM PS therefore exhibits a larger relative suppression on large scales than on smaller scales.

The impact of the suppression on the VID signal is more subtle. As $\sigma^{\chi b}_{n}$ increases, the suppression extends to progressively lower $k$, reducing the abundance of increasingly massive halos that can host galaxies with strong line emissions. This changes the distribution of intensities across voxels, leading to a suppression of the PDF at intermediate and higher brightness temperatures. Since the PDF is normalized, this redistribution of probability enhances $\mathcal{P}(T)$ at lower brightness temperatures, shifting the mean of the distribution toward lower temperatures, as seen in the right panel of Fig.~\ref{fig:LIM_observables_ERA_DMB_10MeV_30GHz}. Subtracting the mean temperature introduces an additional shift in the zero point of the temperature fluctuations, resulting in a corresponding shift of the VID.

 \section{Experimental Setup and Survey Specifications}
\label{sec:app_Exp_Setup_Surv_Spec}
CO is a particularly advantageous tracer for LIM, as it is commonly found in molecular hydrogen clouds, the primary fuel for star formation \cite{Bernal:2022jap, COMAP:2021nrp,Bernal:2022jap,Cleary:2021dsp,Li_2016}, providing a more direct connection to the underlying galaxy population. In contrast, the 21cm line primarily traces neutral hydrogen in the IGM during the epoch of reionization and is indirectly connected to the galaxies hosting the ionizing sources. Molecular clouds are composed primarily of molecular hydrogen ($\mathrm{H}_2$), but
cold $\mathrm{H}_2$ is difficult to observe directly because, as a symmetric molecule, it has no permanent electric dipole moment (EDM); its lowest rotational
transitions are therefore weak quadrupole transitions requiring relatively high excitation temperatures \cite{Li_2016,Bernal:2022jap}. CO, the second
most abundant molecule after $\mathrm{H}_2$, has a permanent EDM and low-lying rotational transitions that are readily excited in molecular
clouds \cite{Bernal:2022jap,COMAP:2021nrp}. In addition, CO observations are less affected by severe foregrounds, although CO surveys are subject to their own line-interloper contamination \cite{Cleary:2021dsp}. Moreover, established empirical and simulation-based relations between CO luminosity, molecular gas, star formation, and halo mass connect the CO intensity field to the underlying galaxy and halo populations, making CO a well-motivated tracer for LIM. 

This motivated the development of the COMAP \cite{Cleary:2021dsp,COMAP:2021qdn,COMAP:2021pxy,COMAP:2021sqw,COMAP:2021lae,COMAP:2021nrp,Rennie_2022}, with the Pathfinder experiment targeting CO(1–0) emission at $z\sim2.4–3.4$ using observations around 30 GHz with a 10.4 m antenna at the Owens Valley Radio Observatory (OVRO). Building on this foundation, next-generation extensions \cite{COMAP:2021nrp} aim to probe higher redshifts and improve sensitivity.

\begin{figure}
    \centering
    \includegraphics[width=1.0\linewidth]{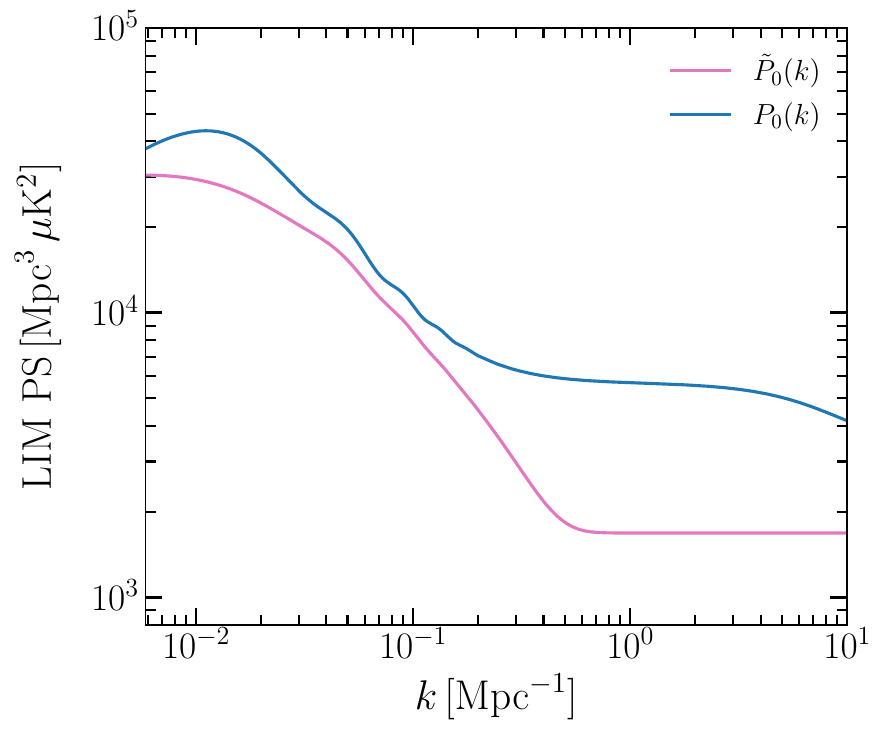}
    \caption{Effect of the experimental window functions on the total LIM monopole PS for $\Lambda$CDM in the COMAP-ERA 30 GHz configuration. The \textbf{pink} curve includes the effects of the finite instrumental resolution and survey volume through the resolution and volume window functions, $W_{\rm res}$ and $W_{\rm vol}$, respectively, while the \textbf{blue} curve shows the unsmoothed prediction. The instrumental resolution suppresses small-scale (high-$k$) modes, while the finite survey volume restricts the accessible large-scale (low-$k$) modes.}
    \label{fig:LIM_PS_ERA_30GHz_smoothing_Test}
\end{figure}
The COMAP-EoR experiment extends the existing Pathfinder by incorporating two additional 10.4 m dishes operating at 30 GHz, along with a new 18 m dish equipped with a 16 GHz receiver \cite{COMAP:2021nrp}. The 30 GHz array primarily probes CO(1–0) (115.27 GHz) emission at $z\sim3$ and CO(2–1) (230.538 GHz) at $z\sim6$, while the 16 GHz receiver accesses CO(1–0) at higher redshifts $z\sim6$. The 16 GHz instrument employs dual-polarization feeds, enhancing sensitivity at high redshift. 

The observing strategy assumes a total campaign comprising 12 years with the Pathfinder, 7 years with the 16 GHz instrument, and 5 years with each of the new 30 GHz dishes, corresponding to approximately 29,000 dish-hours per field at 30 GHz and  7000 dish-hours per field at 16 GHz \cite{COMAP:2021nrp}. For the next-generation COMAP-ERA, the array is envisioned to expand to 10 dishes at each frequency, operating for an additional five years, yielding roughly 110,000 dish-hours at 30 GHz and 57000 dish-hours at 16 GHz \cite{COMAP:2021nrp}. The effective observing time is rescaled to account for the improved sensitivity of the newer receivers \cite{COMAP:2021nrp}. For both frequency bands, we adopt a bandwidth of $\Delta\nu=8$~GHz and, for simplicity, neglect the evolution of the LIM signal across each bandwidth. Since the CO(2–1) signal is subdominant compared to CO(1–0), we neglect its contribution assuming no interloper contamination in our analysis \cite{Sabla:2024dxz}.

The finite instrumental resolution and limited survey volume in LIM experiments modify the observed PS relative to the ideal theoretical prediction in Eq.~\ref{eq:tot_LIM_PS}, primarily by restricting the range of accessible spatial scales. To account for these effects, we follow \cite{Sabla:2024dxz,Sato-Polito:2022fkd,COMAP:2021nrp} and incorporate window functions that encode the impact of resolution and survey geometry. The observed PS can then be written as,
\begin{equation}
\label{eq:obs_PS}
    \tilde{P}_{\rm tot}(k,\mu, z)=\int \frac{d^3q}{(2\pi)^3} W^2_{\rm vol}(\bold{q})W^2_{\rm res}(\bold{q}-\bold{k})P_{\rm tot}(\bold{q}-\bold{k}, \mu, z)\,,
\end{equation}
where $P_{\rm tot}$ is defined in Eq.~\eqref{eq:tot_LIM_PS}, and $W_{\rm vol}$, $W_{\rm res}$ describe the effects of the finite survey volume and voxel resolution, respectively, and and $\tilde{P}(k, \mu, z)$ denotes the smoothed LIM PS that we observe. The instrumental frequency and angular resolutions introduce smoothing along the l.o.s and transverse directions are, 
\begin{equation}
    \sigma_\parallel = \frac{c\,\delta\nu(1+z)}{H(z)\,\nu_{\rm obs}}\,, 
\qquad 
\sigma_\perp = \frac{D_{\rm M}(z)\,\theta_{\rm FWHM}}{\sqrt{8\,\rm log\,2}}\,,
\end{equation}
respectively, where $\nu_{\rm obs}$ is the observed frequency, $D_{\rm M}(z)$ is the comoving angular diameter distance, and $\theta_{\rm FWHM}$ is the angular resolution, taken to be $4.5'$ and $3.7'$ for 30 and 16 GHz, respectively \cite{Sabla:2024dxz,COMAP:2021nrp}. We adopt a spectral resolution $\delta\nu =31.25\, \rm MHz$, consistent with the COMAP Pathfinder analysis \cite{Sabla:2024dxz}. These effects are modeled in Fourier space using a Gaussian window function \cite{Sabla:2024dxz,Bernal:2023ovz,Bernal:2022jap,Bernal:2019jdo}, 
\begin{equation}
W_{\mathrm{res}}(k,\mu) = \exp\left[-k^2\left(\sigma_{\parallel}^2 \mu^2 + \sigma_{\perp}^2 (1 - \mu^2)\right)\right]\,.
\label{eq:window_res}
\end{equation}
In contrast, the survey window $W_{\rm vol}$ arises from the finite spatial extent of the observed field, corresponding to a truncation in real space and hence a convolution in Fourier space. The survey volume is approximated as a cylinder aligned along the l.o.s., with radial extent 
$L_{\parallel} = c\,\Delta\nu (1+z)/H(z)\,\nu_{\mathrm{obs}}$, and transverse radius $R_{\perp} = D_{\rm M}(z)\,\sqrt{\Omega_{\mathrm{field}}/\pi}$, where $\Omega_{\rm field}$ is the survey area.
Assuming uniform observational coverage, the corresponding window function is taken to be a \textit{top-hat}. In Fourier space, this yields \cite{Sabla:2024dxz}
\begin{equation}
    W_{\mathrm{vol}}(k_{\parallel}, k_{\perp}) =
\frac{2\pi R^2_{\perp} L_{\parallel}}{V_{\mathrm{field}}}
\, \frac{J_1(k_{\perp} R_{\perp})}{k_{\perp} R_{\perp}}
\, \mathrm{sinc}\left(\frac{k_{\parallel} L_{\parallel}}{2}\right)\,,
\end{equation}
where $V_{\mathrm{field}} = \pi R_{\perp}^2 L_{\parallel}$, $k_{\parallel}=k\mu$, 
$k_{\perp}=k
\sqrt{(1-\mu^2)}$, and $J_1$ denotes the Bessel function of the first kind. Fig.~\ref{fig:LIM_PS_ERA_30GHz_smoothing_Test} illustrates the combined impact of the instrumental-resolution and survey-volume window functions on the total LIM monopole PS for $\Lambda$CDM in the COMAP-ERA 30 GHz configuration. The inclusion of these window functions suppresses the power on small scales due to the finite instrumental resolution, while the finite survey volume limits the contribution from large-scale modes. 
\begin{figure}
    \centering
    \includegraphics[width=1.0\linewidth]{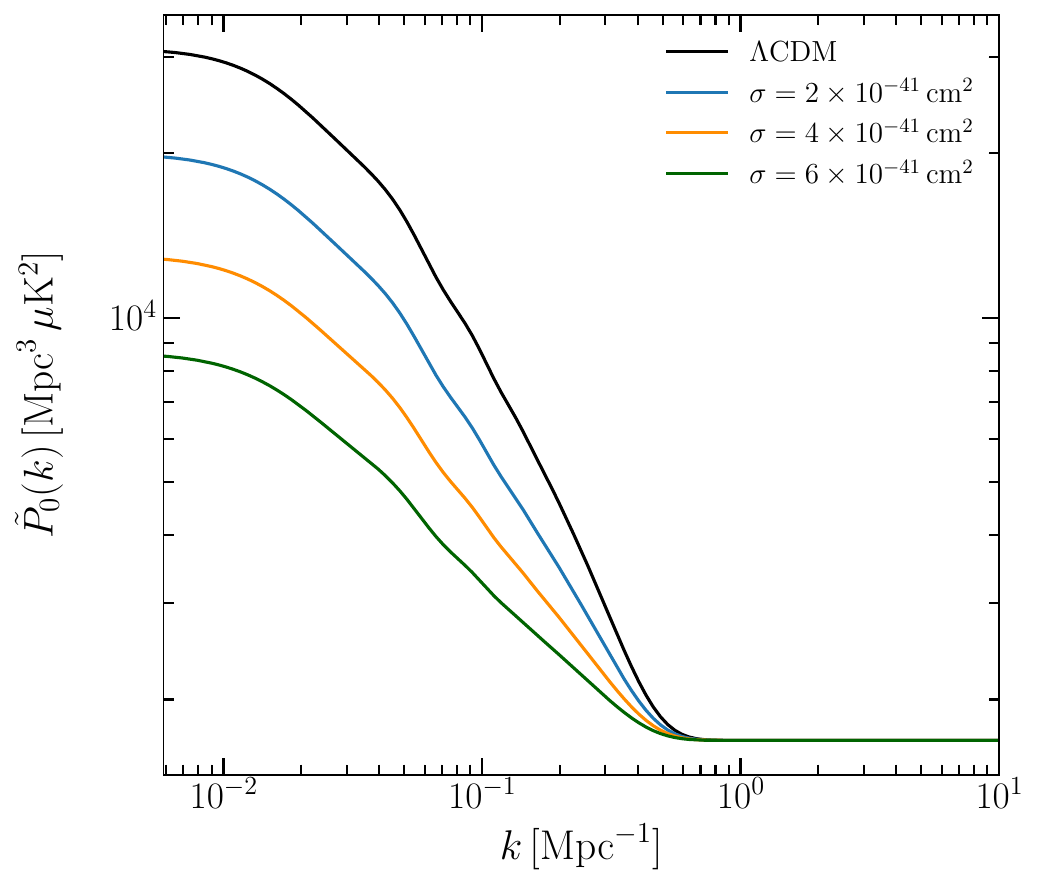}
    \caption{Total monopole PS, $\tilde{P}_0(k)$, after convolution
with the instrumental-resolution and survey-volume window functions for $\Lambda$CDM and representative DM-$p$  interaction scenarios in the case of COMAP-ERA 30 GHz configuration. The \textbf{black} curve denotes the $\Lambda$CDM prediction, while the colored curves correspond to different DM-$p$ scattering
cross sections for $n=-4$ and $m_{\chi}=10\,\rm MeV$.}
    \label{fig:LIM_PS_DMB_ERA_30_GHz_Smooth}
\end{figure}
Fig.~\ref{fig:LIM_PS_DMB_ERA_30_GHz_Smooth} shows the corresponding window-convolved PS for $\Lambda$CDM and representative DM-$p$ interaction scenarios for $n=-4$ and $m_{\chi}=10\, \rm MeV$. The interactions suppress the LIM PS relative to $\Lambda$CDM, with the suppression increasing with the interaction strength.
\begin{figure*}
	\begin{center}
		\includegraphics[width=0.99\columnwidth]{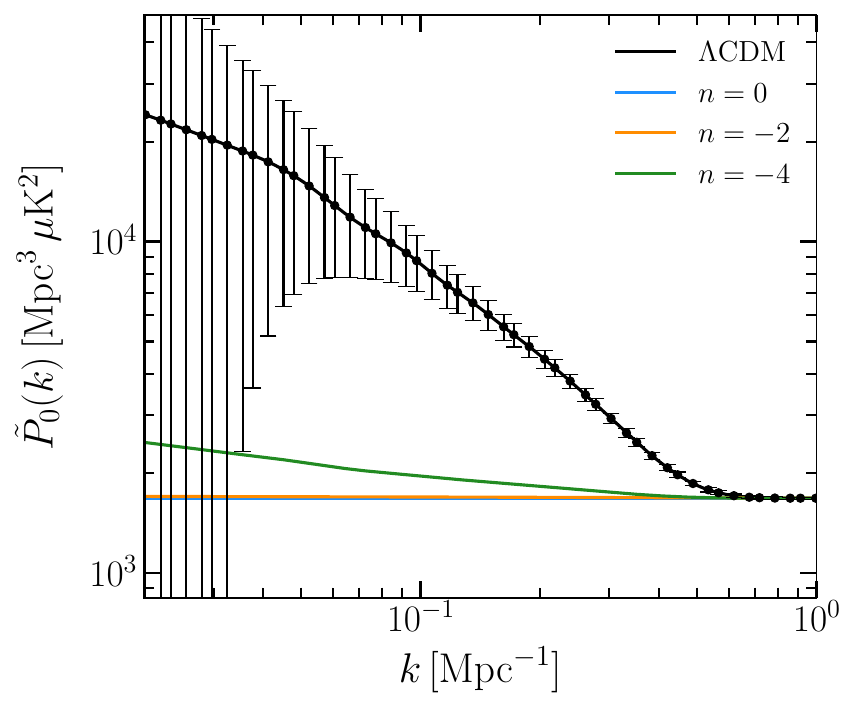}~~
		\includegraphics[width=0.99\columnwidth]{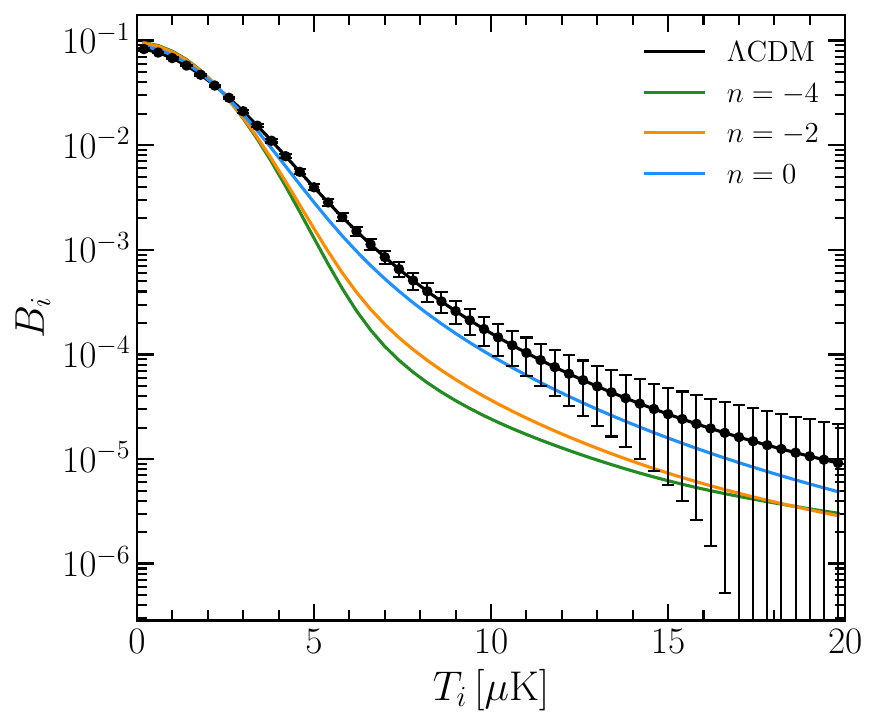}~~\\
		\caption{
		Comparison of the LIM observables for the $\Lambda$CDM model and DM-$p$ interaction models with $n=0,-2,-4$ for the COMAP-ERA 30 GHz configuration. The \textbf{black} points and two-sigma error bars show the fiducial $\Lambda$CDM predictions and their expected error, respectively~\cite{Sabla:2024dxz}. The colored curves show the predictions for the corresponding interaction models evaluated at their projected $95\%$ C.L. upper limits on $\sigma^{\chi p}_{n}$. \textbf{\textit{Left:}} The monopole PS after convolution with the instrumental resolution and survey-volume window functions. \textbf{\textit{Right:}} The corresponding VID.
		}
		\label{fig:Consist_Check_ERA_10MeV_30GHz_alln_v30}
	\end{center}
\end{figure*}

 Instrumental noise arising from practical limitations of the experiment introduces a minimum noise level in the observed brightness-temperature maps. This noise sets a floor in the measurements and contributes a constant component to the PS. Considering only the autocorrelation signal (single-dish experiment) from individual antennas, the variance of the instrumental noise per voxel for a single antenna can be written as \cite{Bernal:2019jdo,Bernal:2022jap,Bull:2014rha}, \begin{equation}
 \sigma^2_{\rm N}=\frac{T^{2}_{\rm sys}}{N_{\rm feeds}N_{\rm pol}\delta\nu_{\rm FWHM} t_{\rm pix}}
\end{equation}
where $T_{\rm sys}$ represents the total system temperature of the instrument which is assigned to be 44 K for 30 GHz and for 16 GHz it is 22 K \cite{COMAP:2021nrp}, $N_{\rm feeds}$ is the number of detectors per antenna which is 19 for both the instruments, and $N_{\rm pol}=1,2$ indicates the number of polarization states measured by the receivers. The quantity $t_{\rm pix}$ corresponds to the effective integration time per pixel, while $\delta \nu _{\rm FWHM} =
\delta \nu \sqrt{8\,\rm log2}$ \cite{Sabla:2024dxz}. In more detailed treatments, the noise level per voxel can additionally be allowed to vary with observing frequency \cite{Chung:2022lpr}.
    Assuming a spatialy uniform and Gaussian noise field the noise PS can be written as \cite{COMAP:2021nrp,Libanore:2022ntl}, \begin{equation}
        P_{\rm noise}=\sigma^2_{\rm N} V_{\rm vox}\,,
    \end{equation}
where $V_{\rm vox}$ is the volume of a voxel.

\section{Constraints  on DM-baryon scattering from COMAP-EoR and COMAP-ERA Forecast}
\begin{figure*}
	\begin{center}
		\includegraphics[width=0.99\columnwidth]{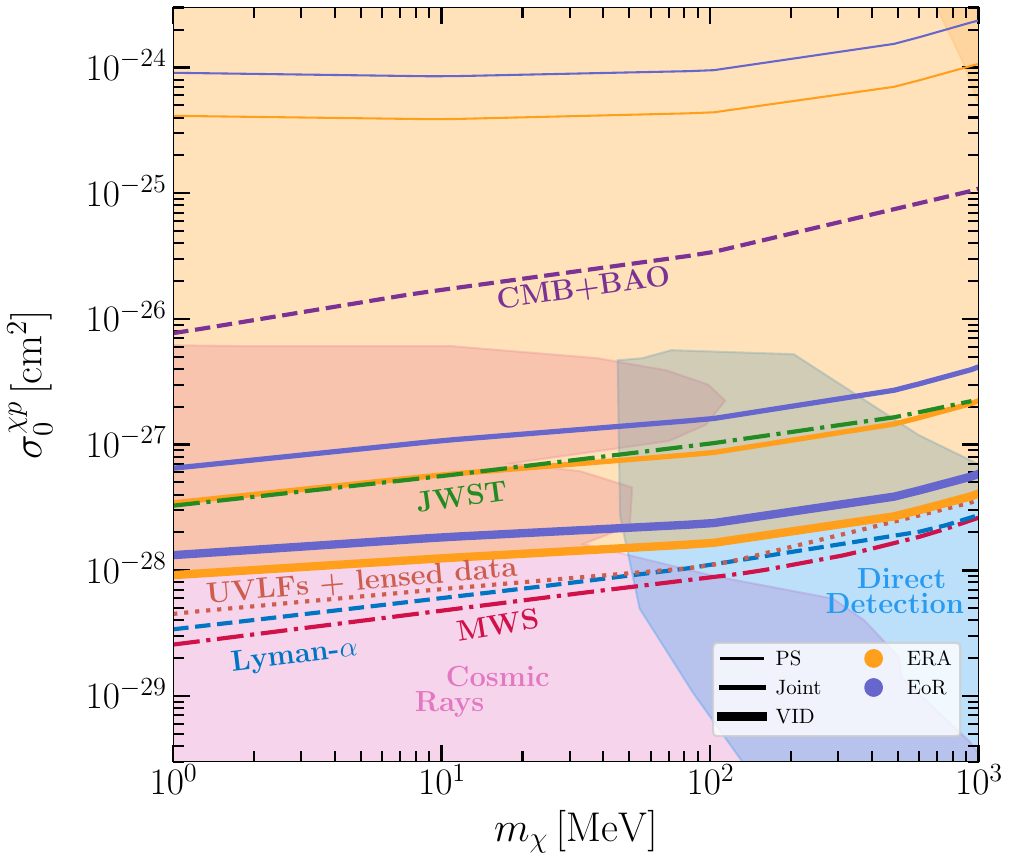}~~
		\includegraphics[width=0.99\columnwidth]{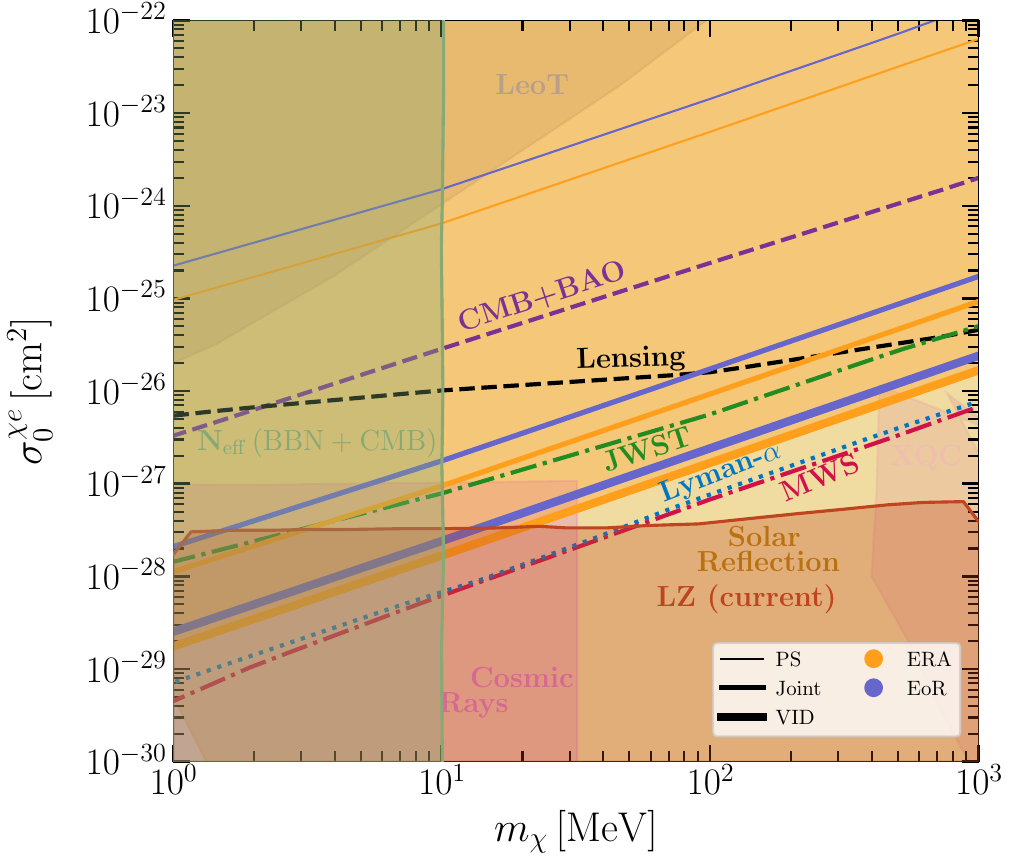}~~\\
		\caption{
		Projected $95\%$ C.L. upper limits on the velocity-independent DM-$b$ scattering cross sections. \textbf{Blue (orange)} curves correspond to the COMAP-EoR (COMAP-ERA) survey configuration, while the \textbf{thin}, \textbf{intermediate-thickness}, and \textbf{thick} curves denote forecasts from the PS, \textit{joint} PS+VID, and VID analyses, respectively. The orange shaded region is disfavored by our \textit{combined} VID analysis for the COMAP-ERA configuration. The other curves show existing cosmological and astrophysical constraints~\cite{Buen-Abad:2021mvc, Nguyen:2021cnb}. The purple dashed line shows the constraint from CMB+BAO. The dash-dotted magenta, blue, and green lines show constraints from MW satellites, the Lyman-$\alpha$ forest, and JWST UVLF spectroscopic data~\cite{Das:2025bnr}, respectively. \textbf{\textit{Left:}} Constraints on the DM-$p$ interaction cross section $\sigma^{\chi p}_0$. The pink shaded region is ruled out by cosmic-ray-boosted DM observations~\cite{Yin:2018yjn}. The blue shaded region is excluded by various direct-detection experiments~\cite{CRESST:2019jnq, CRESST:2019axx, CRESST:2017ues, EDELWEISS:2019vjv, XENON:2017vdw}. The dashed rosy-copper line shows constraints from the HST UVLF+lensed data analysis~\cite{Lazare:2025gha}. \textbf{\textit{Right:}} Constraints on the DM-$e$ interaction cross section $\sigma^{\chi e}_0$. The pink shaded region is ruled out by cosmic-ray observations~\cite{Ema:2018bih}. The yellow shaded region shows constraints from solar reflection~\cite{An:2017ojc, Emken:2021lgc}. The beige-brown shaded region is excluded by gas cooling in the Leo-T dwarf galaxy. The olive-green shaded region shows constraints from BBN+CMB. The pale-peach shaded region is ruled out by XQC. The dark-pink shaded region represents constraints from the LUX-ZEPLIN (LZ) Collaboration~\cite{Maity:2022exk}. The black dashed line shows the constraints obtained from weak lensing~\cite{Zhang:2024mmg}.
		}
\label{fig:n0_DMB_DMe_constraints}
	\end{center}
\end{figure*}

\begin{figure*}
	\begin{center}
		\includegraphics[width=0.99\columnwidth]{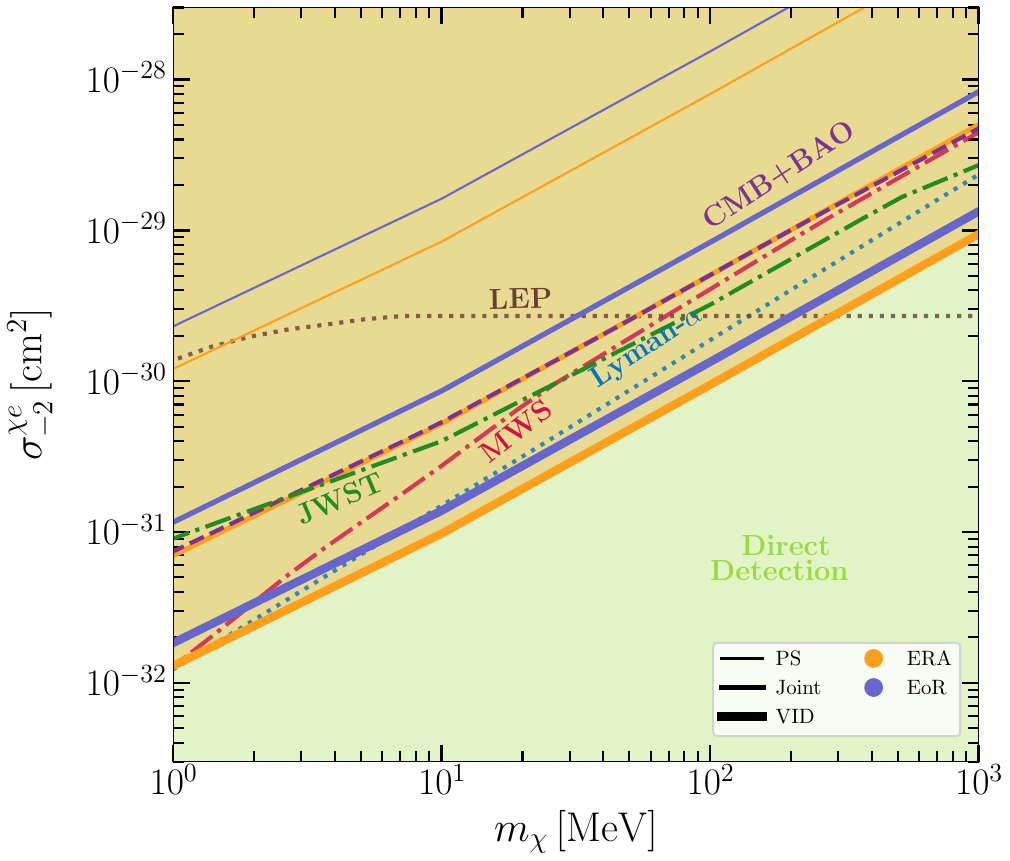}~~
		\includegraphics[width=0.99\columnwidth]{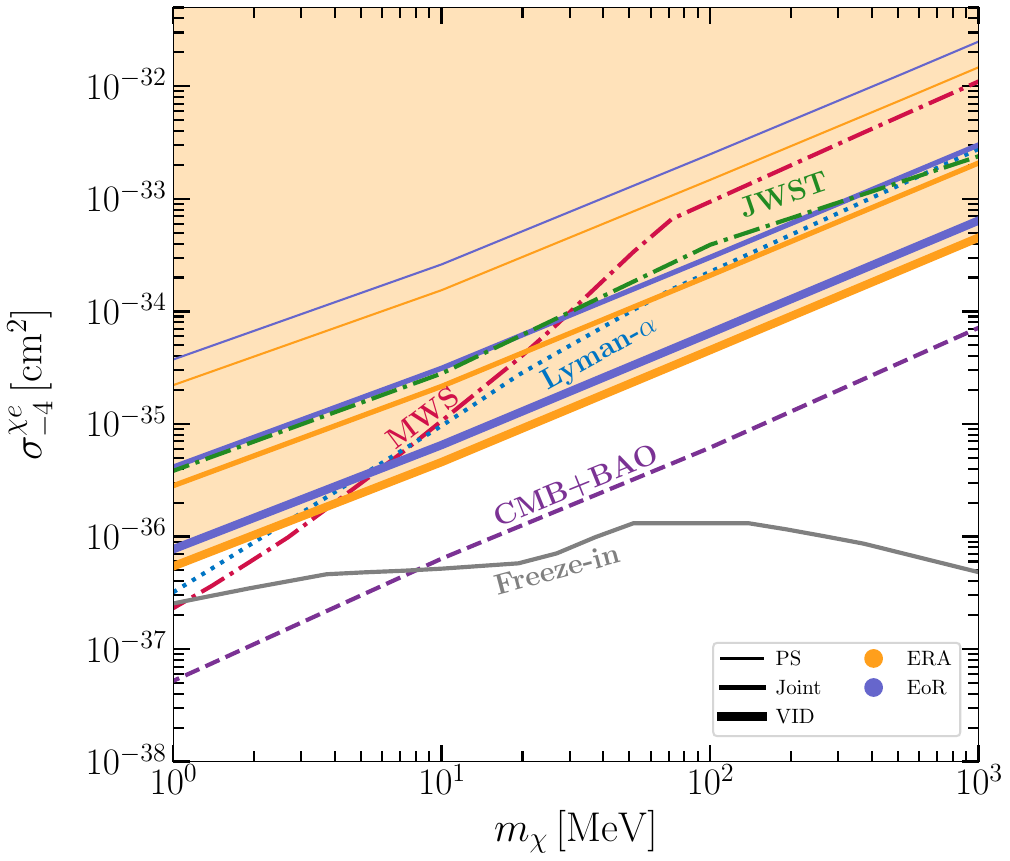}~~\\
		\caption{
		Projected 95\% C.L. upper limits on the normalization DM-$e$ scattering cross sections. \textbf{Blue (orange)} curves correspond to the COMAP-EoR (COMAP-ERA) survey configuration, while the \textbf{thin}, \textbf{intermediate-thickness}, and \textbf{thick} curves denote forecasts from the PS, \textit{joint} PS+VID, and VID analyses, respectively. The orange shaded region is disfavored by our \textit{combined} VID analysis for the COMAP-ERA configuration. The other curves show existing cosmological and astrophysical constraints~\cite{Buen-Abad:2021mvc, Nguyen:2021cnb}. The purple dashed line shows the constraint from CMB+BAO. The dash-dotted red, blue, and green lines show constraints from MW subhalos, the Lyman-$\alpha$ forest, and JWST UVLF spectroscopic data~\cite{Das:2025bnr}, respectively. \textbf{\textit{Left:}} Constraints on the electric dipole-moment-like interaction cross section $\sigma^{\chi e}_{-2}$. The lime-green shaded region is excluded by direct-detection experiments~\cite{XENON:2016jmt, Essig:2012yx, Essig:2017kqs, XENON10:2011prx, DarkSide:2018ppu, SuperCDMS:2018mne, SENSEI:2019ibb, SENSEI:2020dpa}. The brown dotted line represents the upper limits from Ref.~\cite{Fortin:2011hv}. \textbf{\textit{Right:}} Constraints on the Coulomb-like interaction cross section $\sigma^{\chi e}_{-4}$. The gray curve indicates the parameter values for which the observed DM relic abundance is generated entirely through freeze-in production~\cite{Essig:2011nj, Chu:2011be}.
		}
		\label{fig:n-2_n-4_DMe_constraints}
	\end{center}
\end{figure*}

The projected constraints on $\sigma^{\chi p}_{n}$ for $n=0$ and on
$\sigma^{\chi e}_{n}$ for $n=0,-2,-4$ are shown in Figs.~\ref{fig:n0_DMB_DMe_constraints} and
\ref{fig:n-2_n-4_DMe_constraints}, respectively. We compare our results with the existing limits on DM-$b$ interactions for sub-GeV DM masses.

Fig.~\ref{fig:Consist_Check_ERA_10MeV_30GHz_alln_v30} compares the LIM observables for $\Lambda$CDM with those predicted for DM-$p$ interaction models with $n=0,-2,-4$, evaluated at the corresponding
projected $95\%$ C.L. upper limits on $\sigma^{\chi p}_{n}$ for the COMAP-ERA 30 GHz configuration. The PS is shown over $k\simeq0.02$--$1~{\rm Mpc}^{-1}$, while the VID is shown over $T_{i}=0$--$20~\mu{\rm K}$. The black points and two-sigma error bars denote the fiducial $\Lambda$CDM prediction and its expected uncertainties, respectively \cite{Sabla:2024dxz}. The interacting models exhibit a suppression of the LIM PS and a redistribution of the VID relative to $\Lambda$CDM, reflecting the reduced abundance of CO-emitting halos and the associated change in the mean brightness temperature. Interaction models producing deviations larger than the $\Lambda$CDM uncertainties are consequently excluded at the corresponding C.L..

\begin{figure*}
    \centering
    \includegraphics[width=\linewidth]{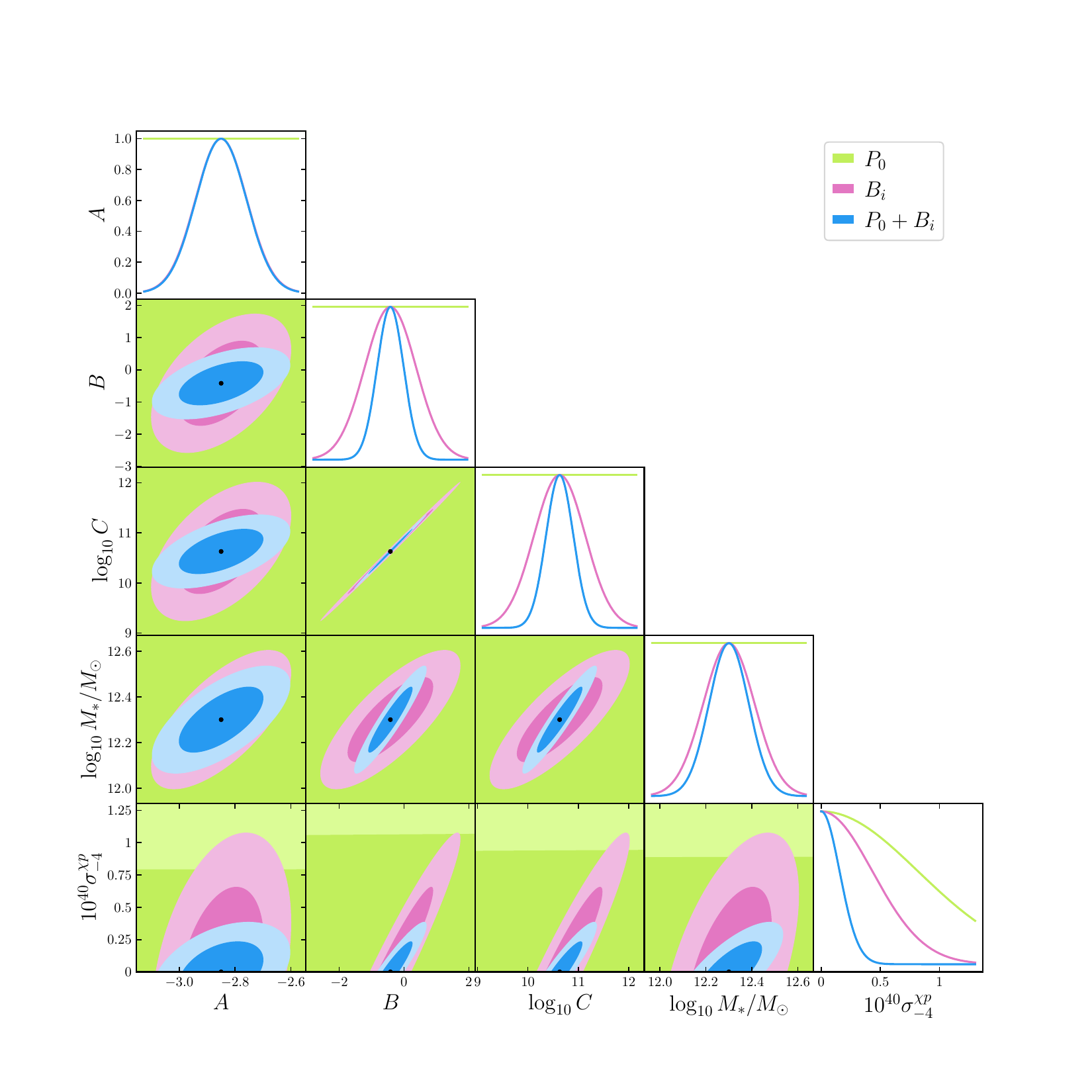}
    \caption{Marginalized one- and two-dimensional $68\%$ and $95\%$ confidence regions for the astrophysical parameters ($A$, $B$, $\rm{log}_{10}$$C$, $\rm{log}_{10}$$(M_*/M_{\odot})$ and the DM–p scattering cross section $\sigma^{\chi p}_{-4}$ for a velocity-dependent $n=-4$ and $m_{\chi}=10\,\rm MeV$, assuming the COMAP-ERA 30 GHz survey configuration. Sensitivities are shown for the LIM PS (\textbf{lime green}), VID (\textbf{pink}), and their \textit{joint} analysis PS+VID (\textbf{blue}).The complementary parameter dependences of the PS and VID help break degeneracies, yielding tighter constraints in the \textit{joint} analysis.}
    \label{fig:Triangle_Plot_ERA_30GHz_DMB}
\end{figure*}

\begin{figure*}
    \centering
    \includegraphics[width=\linewidth]{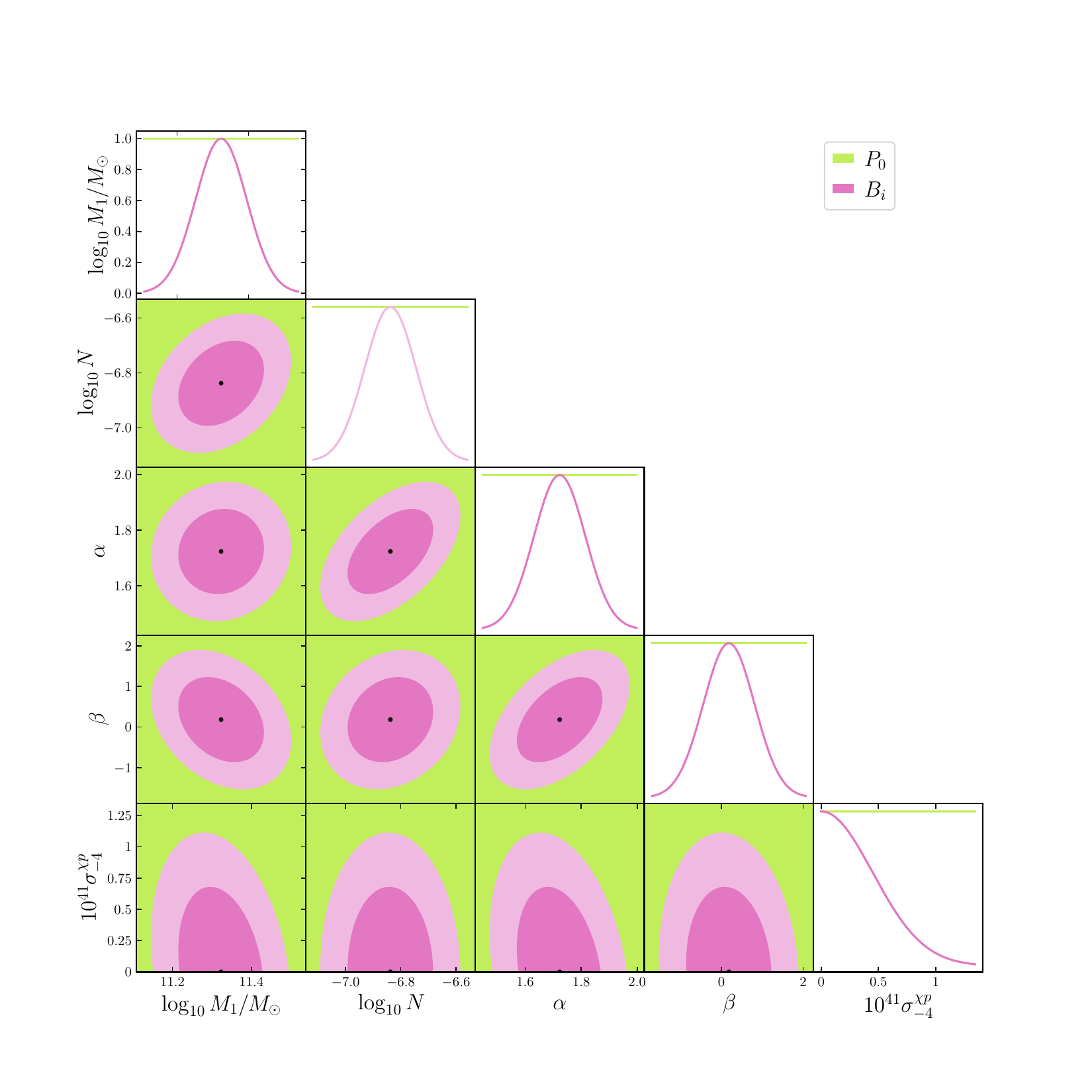}
    \caption{Marginalized one- and two-dimensional $68\%$ and $95\%$ confidence regions for the astrophysical parameters ($\rm{log}_{10}$$M_1$, $\rm{log}_{10}$$N$, $\alpha$, $\beta$) and the DM–p scattering cross section $\sigma^{\chi p}_{-4}$, for a velocity-dependent $n=-4$ and $m_{\chi}=10\,\rm MeV$, assuming the COMAP-ERA 16 GHz survey configuration. Constraints are shown for the LIM PS (\textbf{lime green}), VID (\textbf{pink}). A \textit{joint} analysis is not performed because the covariance matrix required to combine the two observables is unavailable. The VID provides the tightest constraints among the analyses shown.}
    \label{fig:Triangle_Plot_ERA_16GHz_DMB}
\end{figure*}

\begin{figure*}
    \centering
    \includegraphics[width=\linewidth]{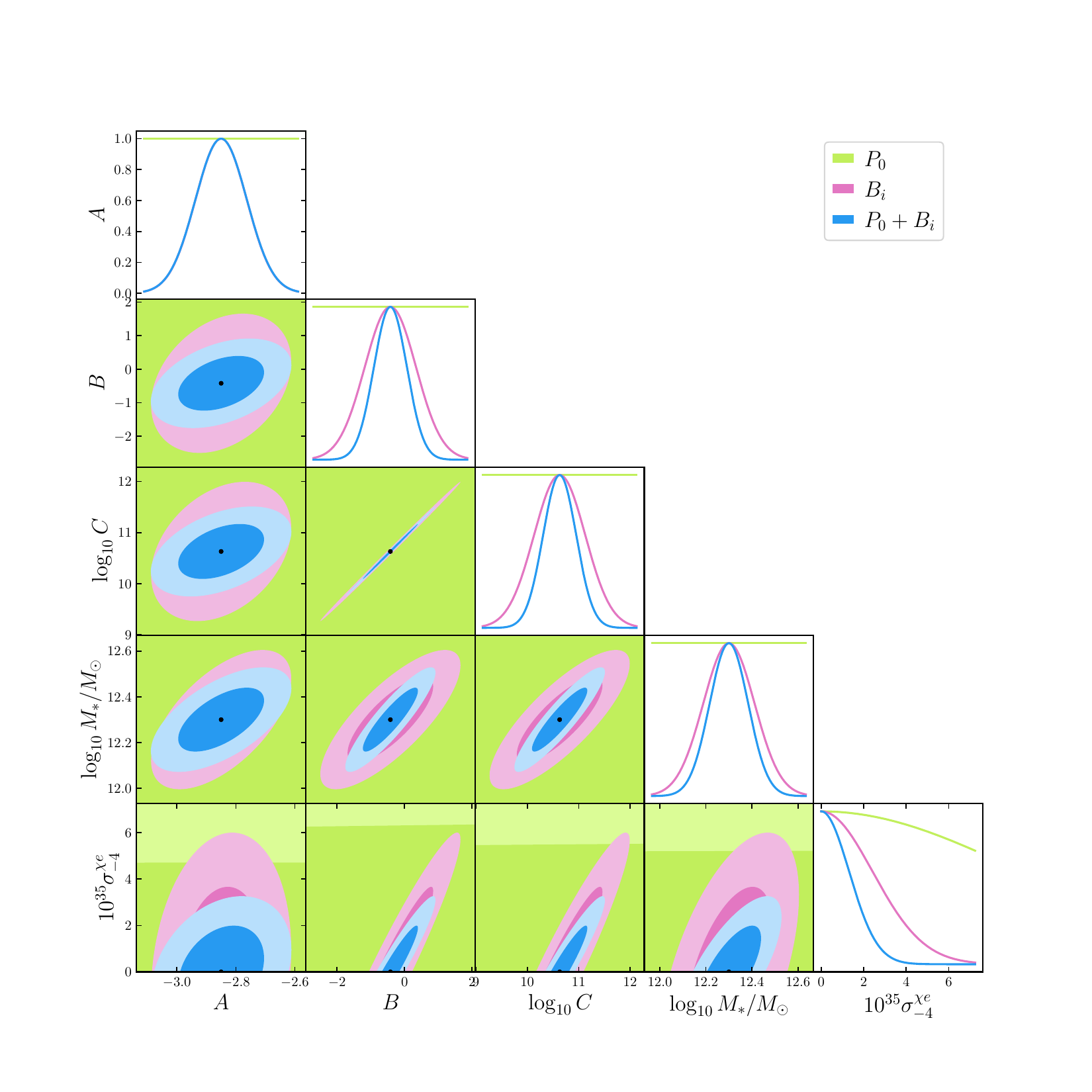}
    \caption{Marginalized one- and two-dimensional $68\%$ and $95\%$ confidence regions for the astrophysical parameters ($A$, $B$, $\rm{log}_{10}$$C$, $\rm{log}_{10}$$(M_*/M_{\odot})$) and the DM–e scattering cross section $\sigma^{\chi e}_{-4}$ for a velocity-dependent $n=-4$ and $m_{\chi}=10\,\rm MeV$, assuming the COMAP-ERA 30 GHz survey configuration. Constraints are shown for the LIM PS (\textbf{lime green}), VID (\textbf{pink}), and their \textit{joint} analysis PS+VID (\textbf{blue}). The \textit{joint} analysis yields the tightest constraints followed by VID and PS.}
    \label{fig:Triangle_Plot_ERA_30GHz_DMe}
\end{figure*}

\begin{figure*}
    \centering
    \includegraphics[width=\linewidth]{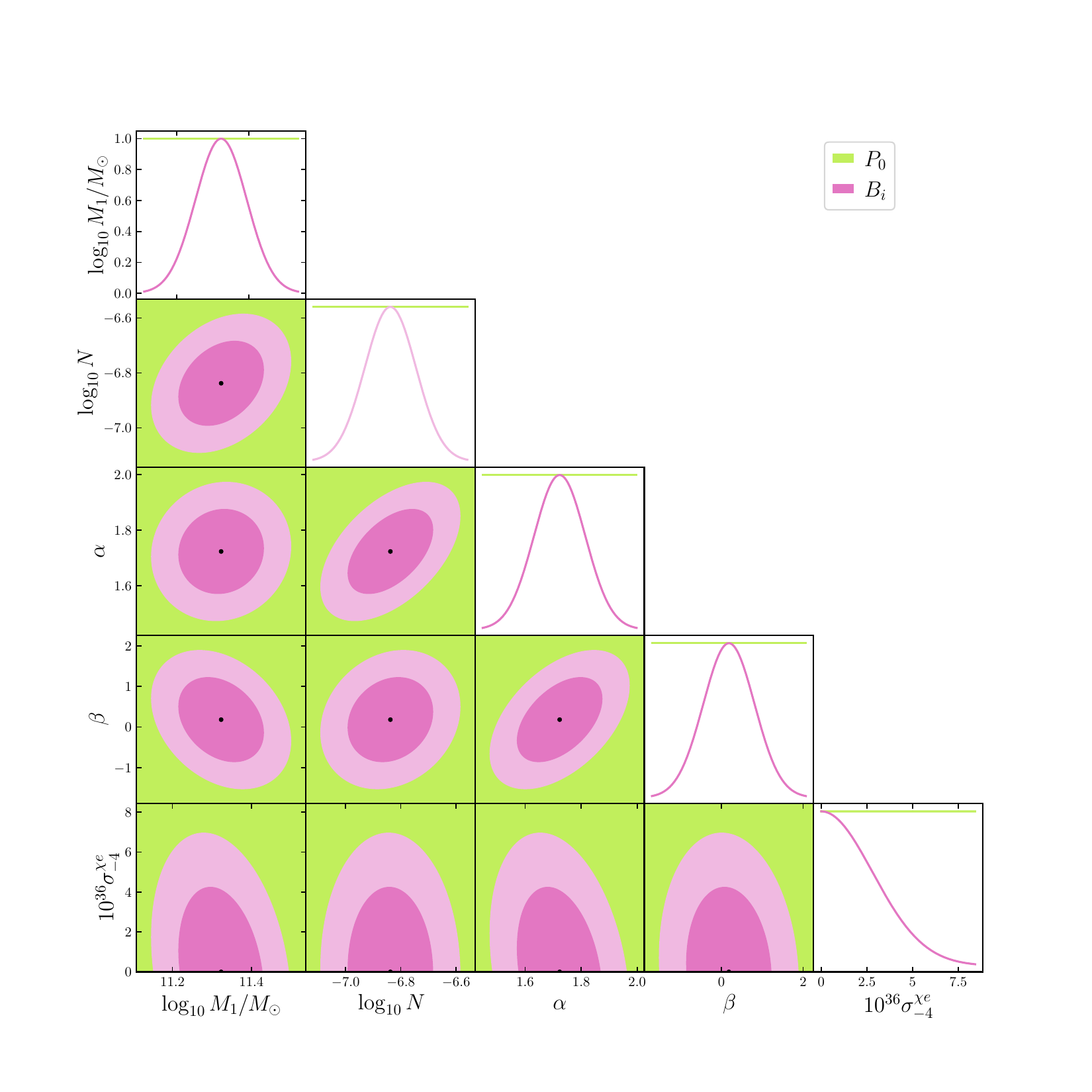}
    \caption{Marginalized one- and two-dimensional $68\%$ and $95\%$ confidence regions for the astrophysical parameters ($\rm{log}_{10}$$M_1$, $\rm{log}_{10}$$N$, $\alpha$, $\beta$) and the DM–$e$ scattering cross section $\sigma^{\chi e}_{-4}$ for a velocity-dependent $n=-4$ and $m_{\chi}=10\,\rm MeV$, assuming the COMAP-ERA 16 GHz survey configuration. Constraints are shown for the LIM PS (\textbf{lime green}), VID (\textbf{pink}). A \textit{joint} analysis is not performed because the covariance matrix required to combine the two observables is unavailable. The VID provides the tightest constraints among the analyses shown.}
    \label{fig:Triangle_Plot_ERA_16GHz_DMe}
\end{figure*}

Figs.~\ref{fig:Triangle_Plot_ERA_30GHz_DMB}--\ref{fig:Triangle_Plot_ERA_16GHz_DMe}
show the marginalized one- and two-dimensional posterior distributions
for the DM-$b$ scattering cross section and the astrophysical
parameters entering the CO luminosity model. Figs.~\ref{fig:Triangle_Plot_ERA_30GHz_DMB} and \ref{fig:Triangle_Plot_ERA_16GHz_DMB} correspond to DM-$p$ interaction for $n=-4$, $m_{\chi}=10\,\rm MeV$ in the COMAP-ERA 30 GHz and 16 GHz configuration respectively, while Figs.~\ref{fig:Triangle_Plot_ERA_30GHz_DMe} and \ref{fig:Triangle_Plot_ERA_16GHz_DMe} show the corresponding constraints for DM-$e$ interactions. For both DM-$p$ and DM-$e$ scattering, the PS, VID, and \textit{joint} analyses exhibit distinct degeneracy directions and posterior widths, reflecting their different dependence on the interaction and astrophysical parameters. The \textit{joint} analysis provides the tightest constraints, followed by the VID and PS analyses, owing to the complementary information contained in the two observables. The \textit{joint} analysis is performed only for the 30 GHz configuration, as the covariance matrix required for the \textit{joint} analysis is not available for the 16 GHz band. 

\section{Constraints on the Dark Matter Electric Dipole Moment}
For a Dirac fermion DM particle, $\chi$, an EDM provides a dimension-five operator to the electromagnetic field. We adopt the effective interaction \cite{Buen-Abad:2021mvc, Chu:2018qrm} \begin{equation}
    \mathcal{L}_{\rm EDM}=\frac{i}{2}d_{\chi}\overline{\chi}\sigma_{\mu \nu}\gamma^5\chi F^{\mu \nu}\,,
\end{equation}
where $d_{\chi}$ is the EDM of DM, $F^{\mu \nu}$ is the electromagnetic field strength tensor, and $\sigma_{\mu \nu}=\frac{i}{2}[\gamma_\mu,\gamma_\nu]$. The interaction mediates DM scattering with charged baryons through photon exchange. 

We consider
\begin{equation}
\chi(k_1)+b(k_2)\rightarrow\chi(k_3)+b(k_4)\,,
\end{equation} where $k_1$ and $k_2$ ($k_3$ and $k_4$) denote the four-momenta of the incoming (outgoing) DM particle and baryon, respectively. The four-momentum transferred in the scattering is $q=k_3-k_1$, with the corresponding Mandelstam variable $t=q^2$. At leading order, the photon propagator gives a $1/t$ enhancement. In the non-relativistic limit, the squared scattering amplitude averaged over initial spins and summed over final spins is \cite{Buen-Abad:2021mvc}
\begin{equation}
    \overline{|\mathcal{M}|^2}\approx\frac{64 \pi \alpha \, d^2_{\chi}m^2_{\chi}m^2_b}{-t}\,,
\end{equation}
where $\alpha$ is the electromagnetic fine structure constant, $m_b$ is the mass of the baryon. The resulting $1/(-t)$ dependence enhances the scattering at small momentum transfer. For proton scattering, the finite electromagnetic structure of the proton can additionally be included through its electric and magnetic form factors \cite{Chu:2018qrm}. 

It is convenient to parameterize the strength of this momentum-dependent interaction by a reference cross section, \begin{equation}
    \overline{\sigma}^{\chi b}=\frac{4\alpha d^2_{\chi}\mu^2_{\chi b}}{q^2_{\rm ref}}\,,
\end{equation}
where $\mu_{\chi b}=m_\chi m_b/(m_\chi+m_b)$ is the reduced mass of the DM-$b$ system and $q_{\rm ref}$ is a fixed reference momentum, introduced to normalize the momentum-dependent EDM interaction. This gives the momentum-dependent cross-section $\sigma^{\chi b}(q)\sim \overline{\sigma}^{\chi b}q^2_{\rm ref}/q^2$.

For cosmological DM–b scattering, however, the relevant quantity is the momentum-transfer cross section, which weights each scattering by the momentum transferred in the collision \cite{Buen-Abad:2021mvc, Dvorkin:2013cea, Nguyen:2021cnb}, \begin{equation}
\sigma^{\chi b}_{T} =
\int d\Omega
(1-\cos\theta)
\frac{d\sigma^{\chi b}}{d\Omega}\,,
\end{equation}
where $\theta$ is the scattering angle in the center-of-mass frame and $d\Omega$ is the differential solid angle.
Using the EDM matrix element above, the vacuum, non-relativistic result is \cite{Buen-Abad:2021mvc}
\begin{equation}
\sigma^{\chi b}_{T} = \frac{2\alpha d_\chi^2}{v^2}
\equiv
\sigma^{\chi b}_{-2}v^{-2}\,,
\end{equation}
where $\sigma^{\chi b}_{-2}=2\alpha d_\chi^2$ and $v$ is the relative velocity between DM and baryon. Thus, EDM-mediated DM–b scattering has the characteristic \(v^{-2}\) dependence, corresponding to \(n=-2\) in the usual parameterization$\sigma^{\chi b}_{T,n}=\sigma^{\chi b}_n v^n$. 

In the baryonic plasma, thermal effects modify the photon propagator through Debye screening. However, the resulting correction is negligible in the regime of interest~\cite{Buen-Abad:2021mvc}. We therefore neglect these thermal effects and use the vacuum momentum-transfer cross section in our analysis. The reference and normalization cross-sections are then related by \cite{Buen-Abad:2021mvc}
\begin{equation}
\overline{\sigma}^{\chi b}_{-2}=\frac{2\mu_{\chi b}^2}{q_{\rm ref}^2}
\sigma^{\chi b}_{-2}\,.
\end{equation}

\begin{figure*}
	\begin{center}
		\includegraphics[width=0.99\columnwidth]{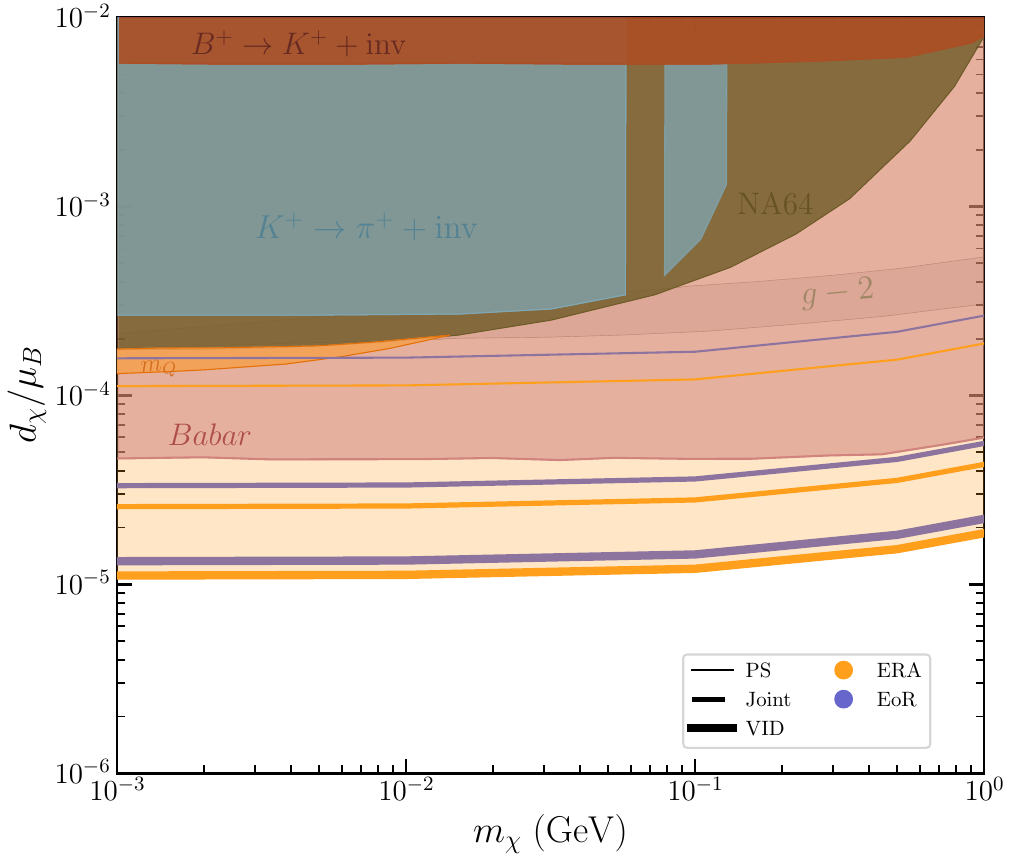}~~
		\includegraphics[width=0.99\columnwidth]{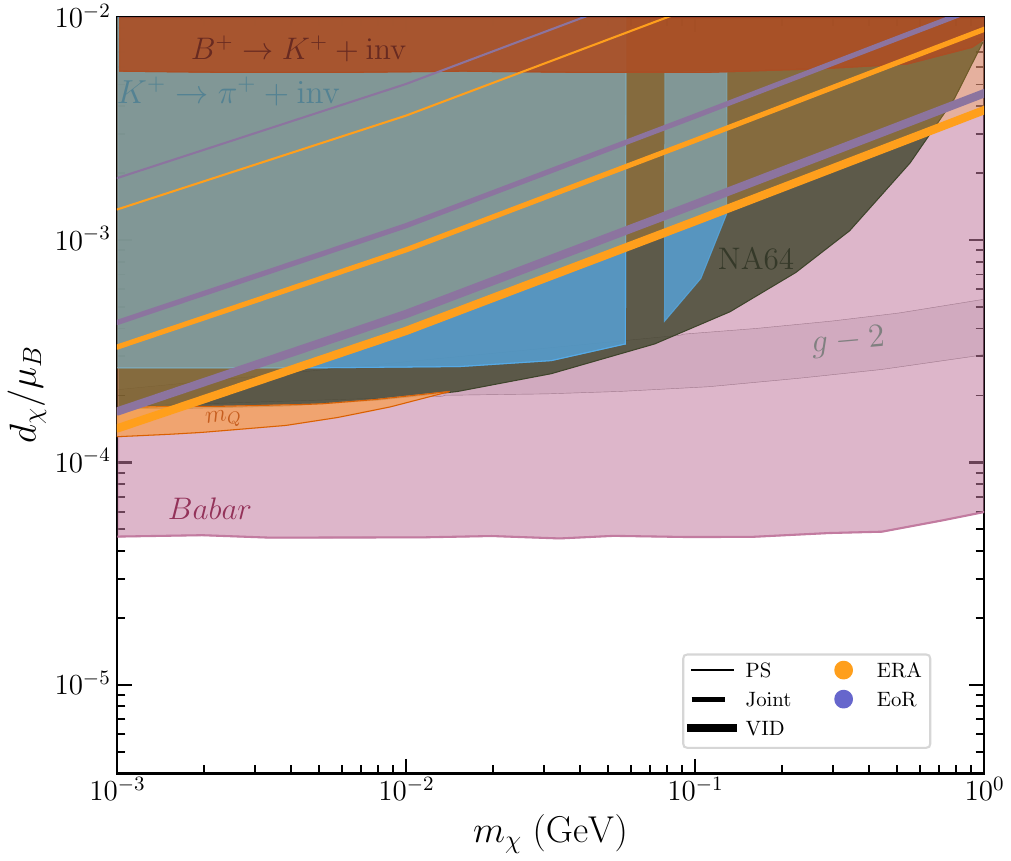}~~\\
		\caption{
		Constraints and projected sensitivities on the DM EDM $d_\chi$ as a function of the DM mass $m_\chi$, for DM-$p$ (\textbf{left}) and DM-$e$ (\textbf{right}) scattering. Existing $90\%$ C.L. laboratory constraints \cite{Chu:2018qrm} are shown together with the projected $90\%$ C.L. sensitivities from LIM with COMAP-EoR (\textbf{blue}) and COMAP-ERA (\textbf{orange}). The \textbf{thin}, \textbf{thick}, and \textbf{intermediate-thickness} curves denote the forecasts from the \textit{combined} PS, \textit{combined} VID, and their \textit{joint} analysis at 30 GHz, respectively.
		}
		\label{fig:d_chi_electron_proton}
	\end{center}
\end{figure*}

For our LIM analysis, we constrain the EDM through the $n=-2$ DM–$b$ interaction. The Fisher analysis gives the sensitivity to the corresponding normalization of the $n = -2$ cross-section $\sigma^{\chi b}_{-2}$, which can be directly translated into a sensitivity on the EDM using $ d_\chi=\sqrt{\sigma^{\chi b}_{-2}/2\alpha}\,.$
We therefore present our LIM forecasts as projected limits on $d_\chi$ as a function of $m_\chi$ in Fig.~\ref{fig:d_chi_electron_proton} for DM-$p$ and DM-$e$ interaction. The figure also shows existing laboratory and rare-meson-decay constraints, together with projected sensitivities from future experiments.


\end{document}